\documentclass[lettersize,journal]{IEEEtran}
\usepackage{amsmath,amsfonts}
\usepackage{algorithmic}
\usepackage{algorithm}
\usepackage{amssymb}
\usepackage{array}
\usepackage{graphicx}
\usepackage{subcaption}
\usepackage{textcomp}
\usepackage{stfloats}
\usepackage{makecell}
\usepackage{url}
\usepackage{upgreek}
\usepackage{verbatim}

\usepackage{colortbl}
\usepackage{varwidth}
\usepackage{microtype}
\usepackage{booktabs}

\usepackage{cite}
\begin{document}

\title{Super-Resolution ISAC Receivers: An MCMC-Based Gridless Sparse Bayesian Learning Approach}

\author{Keying Zhu, \textsl{Graduate Student Member, IEEE}, Xingyu Zhou, \textsl{Graduate Student Member, IEEE},\\ Jie Yang, \textsl{Member, IEEE}, Le Liang, \textsl{Member, IEEE}, and Shi Jin, \textsl{Fellow, IEEE}

\thanks{This paper is an extended version of the conference paper \cite{zhu2025mcmc}.}
}
\markboth{Journal of \LaTeX\ Class Files,~Vol.~14, No.~8, August~2021}%
{Shell \MakeLowercase{\textit{et al.}}: A Sample Article Using IEEEtran.cls for IEEE Journals}

\maketitle

\begin{abstract} 	
Integrated sensing and communication (ISAC) is crucial for low-altitude wireless networks (LAWNs), where the safety-critical demand for high-accuracy sensing creates a trade-off between precision and complexity for conventional methods. 
To address this, we propose a novel gridless sparse Bayesian learning (SBL) framework for joint super-resolution multi-target detection and high-accuracy parameter estimation with manageable computational cost. Our model treats target parameters as continuous variables to bypass the grid limitations of conventional approaches. 
This SBL formulation, however, transforms the estimation task into a challenging high-dimensional inference problem, which we address by developing an enhanced gradient-based Markov chain Monte Carlo algorithm. Our method integrates mini-batch sampling and the Adam optimizer to ensure computational efficiency and rapid convergence.
Finally, we validate the framework's robustness in strong clutter and provide a theoretical benchmark by deriving the corresponding Bayesian Cramér-Rao bound.
Simulation results demonstrate remarkable super-resolution capabilities, successfully resolving multiple targets separated by merely 50$\%$ of the Rayleigh limit in range, 17$\%$ in velocity, and 52$\%$ in angle. At a signal-to-noise ratio of 20 dB, the algorithm achieves a multi-target detection probability exceeding 90$\%$ while concurrently delivering ultra-high accuracy, with root mean square error of 0.07 m, 0.024 m/s, and 0.015° for range, velocity, and angle, respectively. This robust performance, demonstrated against strong clutter, showcases its suitability for practical ISAC-LAWNs applications.
\end{abstract}


\vspace{-2.5mm}
\section{Introduction}
\IEEEPARstart{L}{ow-altitude} wireless networks (LAWNs) are rapidly expanding sixth-generation (6G) systems into a dynamic airspace for applications like UAV logistics and urban air mobility, creating a formidable management challenge \cite{LF_UAVsurvey,yuan_UAV}. This ecosystem fundamentally demands both ultra-reliable, low-latency communication for robust control and high-precision, real-time sensing for critical functions like collision avoidance \cite{control}. Integrated sensing and communication (ISAC) emerges as the key enabling technology, efficiently delivering these dual functions by leveraging shared resources \cite{LF_survey,yangStru}. Modern ISAC systems are increasingly built on cellular networks, whose advanced MIMO and wideband OFDM technologies inherently support simultaneous range-velocity-angle sensing while maintaining reliable communication \cite{ALiusurvey, Azhangsurvey,PMN_cs,Luosurvey}. However, implementing ISAC in practical LAWNs introduces significant sensing challenges. The environment features a diverse mixture of targets, from small, low-RCS aerial objects to large, high-RCS ground vehicles with vastly different kinematics. In this safety-critical context, their high mobility demands both exceptional estimation accuracy and real-time signal processing to ensure continuous situational awareness \cite{UAV_ch,FZY_UAV}.

Several methods have been proposed to overcome these challenges, among which traditional fast Fourier transform (FFT)-based ISAC receivers \cite{FFT1} are widely used. These approaches typically perform parameter estimation via separate delay-Doppler processing and subspace-based angle estimation, but often suffer from inter-dimensional coupling artifacts and degraded performance in multi-target or low signal-to-noise ratio (SNR) scenarios.
To improve estimation accuracy, the three-dimensional FFT-based algorithm in \cite{3DFFT} is proposed to enhance joint range-velocity-angle estimation compared to its two-dimensional counterparts, but it remains fundamentally limited by sampling constraints and high computational complexity. Moreover, these FFT-based approaches typically rely on matched filtering, which inherently bounds the multi-target resolution to the Rayleigh limit \cite{skolnik2008radar}. This resolution limitation poses a critical safety risk in congested LAWNs, as failing to resolve UAVs in dense formations undermines collision avoidance. Consequently, achieving super-resolution beyond the Rayleigh limit is not a mere enhancement but a fundamental requirement for safe operations.
While high-resolution techniques like MUSIC \cite{music1} can surpass this resolution barrier, they often struggle with the complexity of joint multi-parameter estimation and exhibit high sensitivity to noise, failing to strike a balance between super-resolution and robust joint estimation.

To address these limitations, compressed sensing (CS) techniques have been widely adopted, leveraging scene sparsity to recover parameters from fewer measurements \cite{GFF_SBL_svm, QiSBL_ISAC}, which enable joint target detection and high-dimension estimation. 
However, conventional CS techniques are constrained by their reliance on constructing large overcomplete dictionaries over a discretized parameter grid. This on-grid formulation introduces a fundamental trade-off for LAWNs: achieving higher precision requires finer grids, which causes the dictionary size to grow exponentially, leading to impractical computational and storage overhead \cite{c&rprior}. 
More critically, this discretization is a primary source of error known as basis mismatch, which occurs when true target parameters lie between grid points. For LAWNs applications where accuracy is important, this inherent error fundamentally degrades estimation performance and limits the potential for super-resolution.
While off-grid variants \cite{tccn_sbl, nomp} offer iterative refinements, their performance remains constrained by the initial grid and susceptible to local optima, thus providing only a partial remedy.

A gridless approach, which operates in a continuous parameter space, is therefore essential to break this impasse. Such a formulation would inherently eliminate basis mismatch and bypass the need for large dictionaries. While methods like atomic norm minimization \cite{ANM} have explored this direction, they often involve complex semidefinite programming and are sensitive to noise. Among the family of CS techniques, sparse Bayesian learning (SBL) \cite{Tipping2001SparseBL} offers a unique advantage. Its principled probabilistic framework provides the natural foundation for modeling target parameters as continuous random variables. This shift transforms this problem from one of discrete dictionary selection to continuous posterior inference, thereby enabling super-resolution and high-precision parameter estimation.

While this gridless SBL formulation offers enhanced modeling flexibility, it transforms the estimation task into a challenging posterior inference problem. The resulting joint posterior over continuous-valued parameters is typically high-dimensional, non-convex, and analytically intractable, rendering conventional point estimators or variational approximations inadequate for capturing multi-modal structures.
A principled solution for this problem is Markov chain Monte Carlo (MCMC), which constructs a Markov chain whose stationary distribution converges to the target posterior, enabling sample-based estimation via Monte Carlo integration \cite{Hastings1970MonteCS}. To improve efficiency, gradient-based variants such as stochastic gradient Langevin dynamics (SGLD) \cite{SGLD} and gradient-based Metropolis-Hastings (MH) \cite{MH_ramdom} have been introduced, leveraging posterior geometry for better proposal design. Although these methods have shown promise in related fields like MIMO detection and channel estimation \cite{zhou2023gd_detection, zhou2024near}, their direct application to the complex posteriors found in ISAC remains open challenges.
Its high dimensionality and multi-peaks nature causes standard samplers to suffer from slow convergence and instability, making them fundamentally incompatible with the real-time and ultra-accuracy demands of ISAC systems. This highlights the critical need for a more efficient MCMC framework that can make the theoretical advantages of the gridless SBL approach computationally viable.

Beyond these fundamental algorithmic limitations, the practical viability of ISAC receivers is contingent on their robustness to the unique environmental challenges posed by the LAWNs. Since the sensing system is typically ground-based, it is highly susceptible to strong, undesired echoes from static or slow-moving objects, known as clutter. These echoes can severely obscure the much weaker signals reflected from dynamic targets, particularly low RCS UAVs, thereby fundamentally limiting detection sensitivity and estimation accuracy \cite{radarclutter1}. Consequently, developing a robust algorithm that performs well in cluttered environments is a critical prerequisite for the reliable deployment of ISAC in realistic LAWNs scenarios.

This paper thus proposes a novel MCMC-based gridless SBL framework for super-resolution ISAC receivers in LAWNs. The main contributions are summarized as follows:
\begin{itemize}
\item \textbf{A gridless SBL framework for super-resolution}: 
We propose a gridless SBL framework that fundamentally reformulates the ISAC parameter estimation task into a continuous-domain inference problem. This approach inherently eliminates the basis mismatch errors in conventional methods, directly enabling joint super-resolution detection and ultra-high accuracy parameter estimation.
\item \textbf{An efficient and scalable gradient-based MCMC sampler}: We design an efficient MCMC sampler to address the high-dimensional inference challenge introduced by the gridless model. The proposed algorithm integrates mini-batch sampling and the Adam-based preconditioning to ensure the rapid convergence and computational scalability required for practical real-time ISAC applications.
\item \textbf{Theoretical bound and near-optimal performance}: We derive the Bayesian Cramér-Rao bound (BCRB) as a tight theoretical performance limit and demonstrate that our algorithm achieves near-optimal estimation accuracy, closely tracking this bound. 
\end{itemize}

\vspace{-1mm}
\textsl{Notations}: Boldface letters $\mathbf{a}$ and $\mathbf{A}$ denote a vector and a matrix, respectively. $\mathbf{A}^{\text{T}}$, $\mathbf{A}^{\text{H}}$, and $\mathbf{A}^{-1}$ denote the transpose, conjugate transpose, and inverse, respectively. $\mathbf{A} \succeq 0$ means that matrix $\mathbf{A}$ is positive semidefinite. $\otimes$, $\odot$, and $\oslash$ denote the Kronecker product, Hadamard product, and element-wise division, respectively.  
$\nabla_{x}f(x)$ denotes the gradient of $f(\cdot)$ with respect to $x$. 
$\mathbb{E}[\cdot]$ represents the statistical expectation or the expectation operator. $\mathbf{I}$ denotes the identity matrix. $\mathbb{R}$ and $\mathbb{C}$ are the sets of real and complex numbers, respectively.

\section{System Model}
\begin{figure}
\centering
\includegraphics[width=0.95\linewidth]{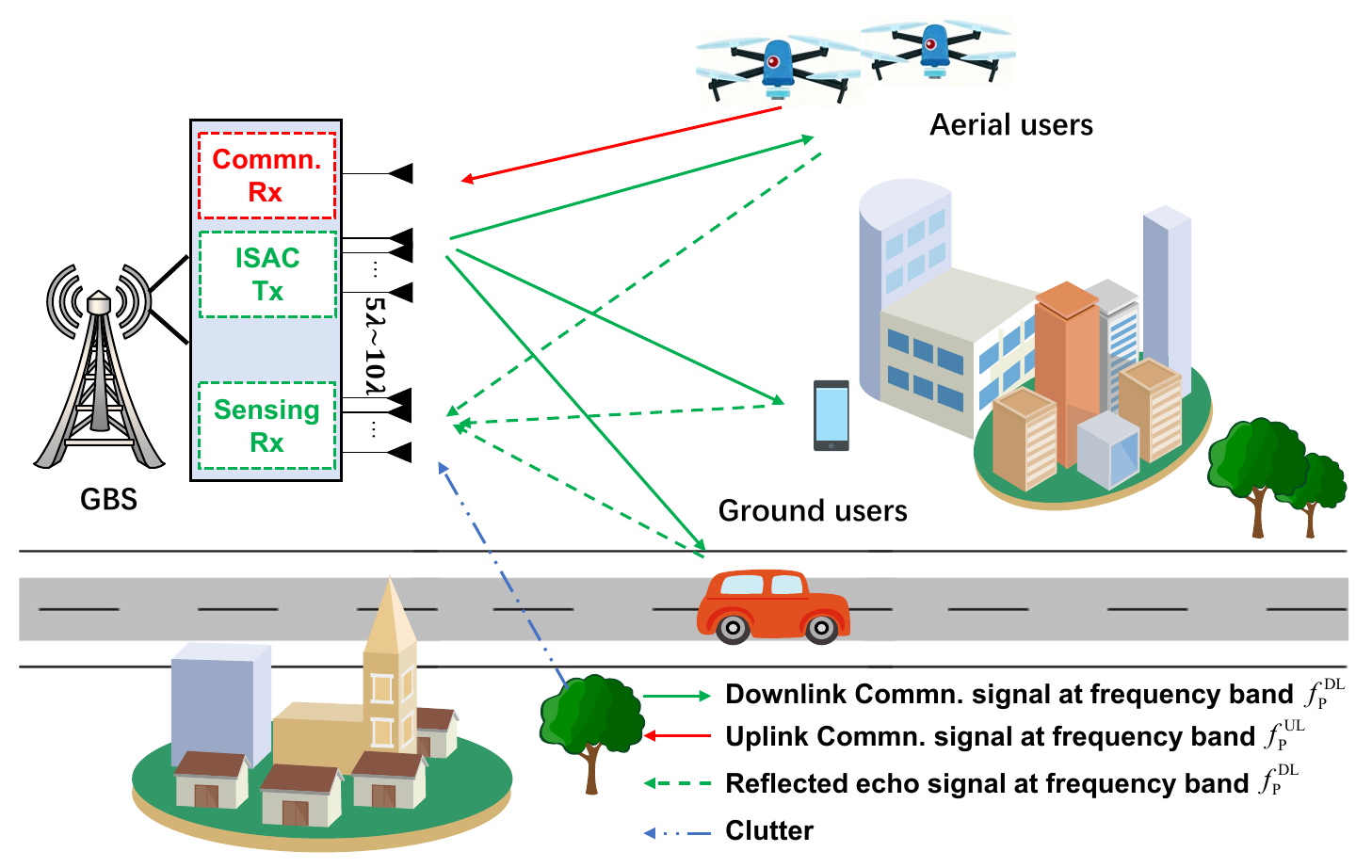}
\captionsetup{font={small}}
\caption{An ISAC-enabled LAWNs scenario with FDD architecture.}
\label{system}
\vspace{-2mm}
\end{figure}

We consider a ground-based mono-static ISAC system designed for surveillance and management of LAWNs, as illustrated in Fig. \ref{system}. A ground base station (GBS) transmits downlink MIMO-OFDM signals serving a dual purpose. First, they provide communication links to a diverse set of users, including $P_{\mathrm{G}}$ ground users and $P_{\mathrm{A}}$ authorized UAVs operating within the airspace. Simultaneously, these signals act as waveforms to illuminate the entire operational environment for sensing.
The dual-functional GBS then captures a complex mixture of echoes reflected by scattering targets in LAWNs. We assume there are $L_{\mathrm{T}}$ potential targets of interest in total, which in this integrated scenario include both highly dynamic low-altitude objects (e.g., cooperative or non-cooperative UAVs) and high-speed ground vehicles. The core task of the GBS receiver is to process these received echo signals to detect the targets and estimate their range-velocity-angle information.
\vspace{-1mm}
\subsection{System Architecture and Transmitted Signal}
To realize a robust ISAC system for LAWNs, which demands continuous and interference-resilient sensing for safety-critical applications, we adopt a frequency division duplexing (FDD) architecture. 
Specifically, the GBS is equipped with a hybrid antenna configuration to support this FDD operation. It features a transmit (Tx) uniform linear array (ULA) with $N$ antenna elements, operating on frequency band $f_p^{\mathrm{DL}}$ to broadcast the downlink signals for both communication and sensing. For reception, a dedicated sensing receive (Sensing Rx) ULA, also with $N$ elements, is co-located and operates on the same band $f_p^{\mathrm{DL}}$ to capture the target echoes. The antenna spacing $d$ for each ULA is set to $\lambda/2$ for optimal spatial resolution, where $\lambda$ is wavelength. Uplink communication, however, is handled by a separate single-antenna communication receiver (Commn. Rx), which is tuned to a non-overlapping frequency band $f_p^{\mathrm{UL}}$. This FDD architecture provides inherent isolation between the different signal types, ensuring that the processing of target echoes at the sensing receiver is immune to interference from uplink signals, thus enabling the system perform uninterrupted surveillance of the airspace and ground environment. 
Additionally, to mitigate self-interference, the downlink ISAC Tx ULA and the Sensing Rx ULA are spatially separated by a distance of approximately $5\lambda$ to $10\lambda$. In practice, the distance between the GBS and the scattering targets is much larger than this separation. Therefore, the small discrepancy in propagation paths caused by this interval is negligible, and a monostatic sensing geometry can be assumed.

The transmitted waveform is a standard MIMO-OFDM signal, leveraging its widespread use in wireless communications and its excellent properties for joint range-velocity-angle sensing. 
The signal consists of $K$ symbols, each containing $M$ subcarriers, and operating at a carrier frequency $f_{\mathrm{c}}$ (within band $f_p^{\rm DL}$) with a subcarrier spacing of $\Delta f$. The total OFDM symbol period $T_{\mathrm{s}}= T_{\mathrm{cp}}+T_{\mathrm{d}}$, where $T_{\mathrm{d}}$ is the data symbol duration equaling $1/\Delta f$, and $T_{\mathrm{cp}}$ is cyclic prefix (CP) duration. To eliminate inter-symbol interference, the CP length is set to be greater than the maximum round-trip delay among all targets. 
The baseband OFDM communication signal $\mathbf{s}(t)\in \mathbb{C}^{N \times 1}$ is then formulated as
\vspace{-1mm}
\begin{equation}
\mathbf{s}(t)=\sum_{k=0}^{K-1}\sum_{m=0}^{M-1}e^{j2 \pi m\Delta f (t-kT_{\mathrm{s}})}\mathbf{x}_{m,k} \mathrm{rect}\Big(\frac{t-kT_{\mathrm{s}}}{T_{\mathrm{s}}}\Big),
\vspace{-1mm}
\end{equation}
where $\mathbf{x}_{m,k}\in \mathbb{C}^N$ is the dual-functional symbol vector for the $m$-th subcarrier and $k$-th OFDM symbol, which is intended for the $N$ Tx antennas. It is designed via precoding to simultaneously serve the $P_{\mathrm{G}}+P_{\mathrm{A}}$ users and illuminate the entire operational environment for sensing. 
The $n$-th element of this vector is the signal destined for the $n$-th Tx antenna.
The function $\mathrm{rect}(t/T)$ depicts a rectangular pulse with a duration of $T$. After baseband generation, $\mathbf{s}(t)$ is up-converted to the radio frequency domain and emitted through the ISAC Tx. 

\vspace{-2mm}
\subsection{Received Echo Signal Model}
After being transmitted by the GBS, the OFDM signals propagate through the low-altitude environment, illuminating both designated communication users and sensing targets. 
The echoes reflected from these $L_{\mathrm{T}}$ high-speed dynamic targets are then captured by Sensing Rx ULA. In our considered LAWN scenario, these echoes are a superposition of reflections from $L_{\mathrm{A}}$ low-altitude aerial targets and $L_{\mathrm{G}}$ ground vehicles, each treated as a point-like source. After down-conversion to baseband, the resulting received aggregate signal vector $\mathbf{y}(t)\in \mathbb{C}^{N\times 1}$ can be expressed as
\vspace{-1mm}
\begin{equation}
\mathbf{y}(t)= \sum_{l=1}^{L_{\mathrm{T}}} b_l e^{j2\pi f_{\mathrm{D},l}t} \mathbf{a}(\theta_{\mathrm{AoA},l})\mathbf{a}^{\mathrm{T}}(\theta_{\mathrm{AoD},l})\mathbf{s}(t-\tau_l) +\mathbf{z}(t),
\label{rx1}
\vspace{-1mm}
\end{equation}
where the summation is over all $L_{\mathrm{T}}=L_{\mathrm{A}}+L_{\mathrm{G}}$ targets. For the $l$-th target, $b_l$ is the complex reflection coefficient, which absorbs the carrier phase $e^{-j2\pi f_{\mathrm{c}} \tau_l}$ and is typically much smaller for UAVs than for ground vehicles. The term $f_{\mathrm{D},l}$ equaling $2v_l/\lambda$ represents the Doppler frequency shift caused by the target's radial velocity $v_l$, and $\tau_l$ defined by $2r_l/c$ indicates the propagation delay, where $r_l$ is the distance between the target and GBS, and $c$ is the speed of light. 
$\theta_{\mathrm{AoA},l}$ and $\theta_{\mathrm{AoD},l}$ represent the angle-of-arrive (AoA) and the angle-of-departure (AoD), respectively. $\mathbf{a}(\theta)$ is the steering vector denoted by
\vspace{-1mm}
\begin{equation}
\mathbf{a}(\theta)=[1,e^{j2\pi \frac{d}{\lambda}\sin\theta},\cdots,e^{j2\pi(N-1)\frac{d}{\lambda}\sin\theta}]^{\mathrm{T}},
\vspace{-1mm}
\end{equation}
where $d$ is antenna spacing. In the considered mono-static GBS system, the number of ISAC Tx and Sensing Rx antennas is equal, and the $l$-th AoD is taken to be identical to the $l$-th AoA, represented by $\theta_l$ for simplicity. $\mathbf{z}(t)$ is the additive white Gaussian noise vector. Assuming a quasi-static channel, the target parameters $\{b_l,\tau_l,f_{\mathrm{D},l},\theta_l\}_{l=1}^{L_{\mathrm{T}}}$ remain constant within a coherent processing interval.

At the receiver, provided that the CP interval is properly chosen, the continuous-time echo signal  $\mathbf{y}(t)$ is cut in sync with the OFDM symbol period, and the delay in the rectangular function can be neglected. 
Considering that $\Delta f$ is much smaller than $f_\mathrm{c}$, the Doppler shift induced on a single
subcarrier is also negligible.
Following these considerations, by removing the CP and applying an $M$-point FFT to each OFDM symbol, the frequency domain echo signal on the $m$-th subcarrier of the $k$-th OFDM symbol $\mathbf{y}_{m,k}\in \mathbb{C}^{N\times1}$ is
\vspace{-1mm}
\begin{equation}
\mathbf{y}_{m,k} \! =\! \sum_{l=1}^{L_{\mathrm{T}}} b_l \underbrace{e^{-j2\pi m \tau_l \Delta f} e^{j2\pi k f_{\mathrm{D},l} T_{\mathrm{s}}}\mathbf{a}(\theta_l) \mathbf{a}^{\mathrm{T}}(\theta_l)\mathbf{x}_{m,k}}_{\boldsymbol{\phi}_{m,k,l}}\! + \mathbf{z}_{m,k},
\label{rx2}
\end{equation}
where $\mathbf{z}_{m,k}$ is the frequency-domain noise vector, whose elements follow an independently and identically distributed (i.i.d.) complex Gaussian distribution $\mathcal{CN}(0,\sigma^2)$. The term $\boldsymbol{\phi}_{m,k,l}\in \mathbb{C}^{N\times 1}$ embeds the range-velocity-angle information of the $l$-th target for this specific time-frequency bin. 
Consequently, the model in (\ref{rx2}) represents the fundamental data block, serving as the foundation for the core detection and estimation problem in our ISAC receivers.

\vspace{-3mm}
\subsection{Problem Formulation via Sparse Representation}
We construct a comprehensive estimation problem from fundamental data blocks by leveraging collective information across the entire time-frequency observation block through vectorization. By stacking the received vectors $\mathbf{y}_{m,k}$ across all $M$ subcarriers and $K$ OFDM symbols, we form a single observation vector $\tilde{\mathbf{y}} \triangleq [\mathbf{y}_{0,0}^\mathrm{T}, \cdots, \mathbf{y}_{0,K-1}^\mathrm{T}, \mathbf{y}_{1,0}^\mathrm{T}, \cdots, \mathbf{y}_{M-1,K-1}^\mathrm{T}]^\mathrm{T} \in \mathbb{C}^{MKN \times 1}$. This leads to a compact linear representation with
\vspace{-1mm}
\begin{equation}
\tilde{\mathbf{y}} = \sum_{l=1}^{L_{\mathrm{T}}} b_l \mathbf{d}(\tau_l, f_{\mathrm{D},l},\theta_{l}) + \tilde{\mathbf{z}},
\label{sparse}
\vspace{-1mm}
\end{equation}
where $\tilde{\mathbf{z}}$ is the aggregated noise vector obtained by stacking all $\mathbf{z}_{m,k}$ in the same order as $\mathbf{y}_{m,k}$, and $\mathbf{d}(\cdot)$ can be treated as the space-time steering vector to the $l$-th target, given by
\begin{equation}
\mathbf{d}(\tau_l, f_{\mathrm{D},l},\theta_{l}) \triangleq \mathrm{vec}([\boldsymbol{\phi}_{0,0,l}, \dots, \boldsymbol{\phi}_{0,K-1,l}, \dots, \boldsymbol{\phi}_{M-1,K-1,l}]).
\end{equation}

While this form offers structural convenience, directly estimating the target information from (\ref{sparse}) remains challenging due to all target parameters that characterize the echoes are unknown, resulting in a highly non-linear and non-convex problem.
The conventional approach to tackle this is to reformulate it within the CS framework, leveraging the inherent sparsity of targets in LAWNs.
The core idea is to transform the non-linear search in a continuous space into a linear problem in a high-dimensional discrete space. 
This is achieved by discretizing the continuous range-velocity-angle parameter space into a fine-grained grid of $G$ candidate tuples, $\{\bar{\tau}_g,\bar{f}_{\mathrm{D},g},\bar{\theta}_g\}_{g=1}^{G}$, where $G$ is chosen to be much larger than the expected number of targets ($G \gg L_{\mathrm{T}}$). 
An overcomplete dictionary $\mathbf{D}_{\mathrm{grid}}\in \mathbb{C}^{MKN \times G}$ is then constructed by stacking the corresponding space-time steering vectors for each grid point
\begin{equation}
\mathbf{D}_{\mathrm{grid}} = [\mathbf{d}(\bar{\tau}_1, \bar{f}_{D,1}, \bar{\theta}_1), \ldots, \mathbf{d}(\bar{\tau}_G, \bar{f}_{\mathrm{D},G}, \bar{\theta}_G)].
\end{equation}
To represent the sparse scene on this grid, a $G$-dimensional vector $\mathbf{b}_{\mathrm{grid}}=[b_{\mathrm{grid},1},\dots,b_{\mathrm{grid},G}]^{\mathrm{T}}$ is introduced, where each entry represents the reflection coefficient at a specific grid point. 
Since there are only $L_{\mathrm{T}}$  true targets, $\mathbf{b}_{\mathrm{grid}}$ is inherently sparse, containing only a few non-zero entries. Crucially, the indices of these non-zero entries directly identify the grid points that best approximate the true target parameters. The original non-linear estimation task thus becomes seeking the sparse vector $\mathbf{b}_{\mathrm{grid}}$ in the following linear system
\begin{equation}
\tilde{\mathbf{y}} =\mathbf{D}_{\mathrm{grid}}\mathbf{b}_{\mathrm{grid}}+\tilde{\mathbf{z}}.
\end{equation}

This conventional grid-based CS approach, however, suffers from fundamental limitations that are severely amplified in the demanding context of LAWNs. 
First, the estimation accuracy is fundamentally constrained by the grid's granularity. 
For LAWN applications requiring precise trajectory tracking, this inherent basis mismatch error limits the system's reliability. 
Second, the diverse kinematics of LAWN targets, combined with the vast surveillance environment, necessitate an exceptionally fine and wide parameter grid to maintain accuracy, causing an exponential growth in dictionary size that renders real-time processing computationally infeasible. 
Finally, for congested air-ground space, the resolution of grid-based methods is fundamentally bounded by the grid spacing, preventing them from achieving the super-resolution required to distinguish closely spaced targets.

These challenges reveal a fundamental trade-off between the high-resolution and practical computation resource in grid-based methods. This impasse motivates a paradigm shift towards a gridless approach. 
We therefore propose a gridless SBL framework, implemented via an efficient MCMC algorithm, to overcome these limitations.

\section{MCMC-Based Gridless SBL Algorithm}
In this section, we develop a novel gridless SBL framework for super-resolution estimation with reduced computational and storage costs. Then, to address the resulting high-dimensional inference challenge, we propose an enhanced gradient-based MCMC algorithm for ultra-high accuracy Bayesian posterior inference. Furthermore, we present a systematic procedure for joint target detection and parameter extraction, which translates the MCMC samples into definitive physical estimates. Finally, we derive the BCRB as a theoretical performance limit and conclude with an analysis of the algorithm complexity.
\vspace{-3mm}
\subsection{A Gridless SBL Framework}
We propose a gridless SBL framework to break the fundamental trade-off between accuracy and complexity inherent in grid-based methods. Instead of discretizing parameter space, our approach embraces the continuous nature of the parameter space by treating both the physical parameters and the complex reflection coefficients as continuous latent random variables to be estimated.
This Bayesian perspective allows us to formulate a gridless model with
\begin{equation}
\tilde{\mathbf{y}} =\mathbf{D}(\boldsymbol{\tau},\boldsymbol{f}_{\mathrm{D}},\boldsymbol{\theta})\mathbf{b}+\tilde{\mathbf{z}},
\label{cs_model}
\end{equation}
where $\mathbf{D}(\cdot) = [\mathbf{d}(\tau_1, \dots), \dots, \mathbf{d}(\tau_Q, \dots)]\in \mathbb{C}^{MKN \times Q}$ is not a fixed, oversized matrix but is continuously parameterized by the unknown parameters $\boldsymbol{\tau},\boldsymbol{f}_{\mathrm{D}},\boldsymbol{\theta} \in \mathbb{R}^{Q\times1}$, and the vector $\mathbf{b} \in \mathbb{C}^{Q\times 1}$ is sparse containing the associated reflection coefficients. 
Among various sparse recovery techniques, SBL offers a unique advantage for realizing this gridless formulation. Its principled probabilistic framework provides the natural foundation for modeling the target parameters $(\boldsymbol{\tau}, \boldsymbol{f}_{\mathrm{D}},\boldsymbol{\theta})$ as continuous random variables with appropriate priors. This paradigm shift transforms the problem from one of discrete dictionary selection to a joint Bayesian inference task over the entire set of unknowns $\{\boldsymbol{\tau}, \boldsymbol{f}_{\mathrm{D}},\boldsymbol{\theta}, \mathbf{b}\}$.

This gridless SBL framework estimates parameters in a continuous domain, fundamentally eliminating the basis mismatch problem and enabling ultra-high estimation resolution and accuracy unattainable with grid-based approaches. Furthermore, unlike the grid-based approach which requires an enormous dictionary size $G$ for fine resolution, the dimension $Q$ only needs to be slightly larger than the maximum expected number of targets. This eliminates the need to construct, store, and operate on a massive dictionary, leading to a substantial reduction in both storage overhead and computational complexity.

In summary, the proposed gridless SBL framework circumvents the fundamental limitations of grid-based methods. To fully realize its benefits, we now proceed to define the complete Bayesian inference model, including the likelihood and prior distributions over the entire set of unknown variables.
As the received signal is corrupted by complex additive Gaussian noise, the likelihood is
\vspace{-1mm}
\begin{equation}
p(\tilde{\mathbf{y}}|\mathbf{b},\boldsymbol{\tau},\boldsymbol{f}_\mathrm{D},\boldsymbol{\theta},\xi)=\mathcal{C}\mathcal{N}(\mathbf{D}\mathbf{b},\xi^{-1}\mathbf{I}),
\vspace{-1mm}
\end{equation}
where we place a Gamma prior on the inverse noise variance $\xi=\sigma^{-2}$ as 
\vspace{-1.5mm}
\begin{equation}
p(\xi)=\text{Gamma}(\xi;\kappa_{\upxi},\chi_{\upxi}) =\frac{{\chi_{\upxi}}^{\kappa_{\upxi}}\xi^{\kappa_{\upxi}-1}e^{(-\chi_{\upxi}\xi)}}{\Gamma(\kappa_{\upxi})},
\label{noise_variance}
\vspace{-1mm}
\end{equation}
where $\Gamma(\cdot)$ is Gamma function, $\kappa_{\upxi}$ and $\chi_{\upxi}$ are hyperparameters. 
By modeling the inverse noise variance $\xi$ as a random variable, our framework can automatically infer the noise level from the data. This adaptability allows the model to mitigate the impact of noise during posterior inference, enhancing the robustness of estimation performance, particularly in low SNR scenarios.

We adopt a two-layer hierarchical prior, to promote sparsity in $\mathbf{b}$. In the first layer, each element $b_{q}$ is modeled as a zero-mean complex Gaussian distribution
\vspace{-1mm}
\begin{equation}
p(\mathbf{b}|\boldsymbol{\rho}) = \prod_{q=1}^{Q}\mathcal{CN}(0,{\rho_{q}}),
\vspace{-1mm}
\end{equation}
where $\boldsymbol{\rho}=[\rho_1,\rho_2,\dots,\rho_Q]^\mathrm{T}$ is a vector of non-negative hyperparameters that govern the variance of each component and thus control the sparsity of $\mathbf{b}$. The second-layer prior distribution of $\boldsymbol{\rho}$ follows a Gamma distribution
\vspace{-1.5mm}
\begin{equation}
p(\boldsymbol{\rho})=\prod_{q=1}^{Q}\text{Gamma}(\rho_q;\kappa_{\uprho},\chi_{\uprho}) =\prod_{q=1}^{Q} \frac{{\chi_{\uprho}}^{\kappa_{\uprho}}{\rho_q}^{\kappa_{\uprho}-1}e^{(-\chi_{\uprho}\rho_q)}}{\Gamma(\kappa_{\uprho})},
\end{equation}
where $\kappa_{\uprho}$ and $\chi_{\uprho}$ are hyperparameters shared by all $\rho_q$ \cite{c&rprior}. When $\rho_q \to 0$, the corresponding prior of $b_q$ converges to a Dirac distribution centered at zero, effectively enforcing $b_q=0$ with high probability. Conversely, when $\rho_q\to \infty$, the prior approaches a broad zero-mean complex Gaussian distribution, allowing $b_q$ to take a nonzero value.

Considering the sensing capability constraints of practical ISAC systems and the maximum unambiguous values of targets, the delay, Doppler shift, and angle are modeled as truncated Gaussian distributions within bounded intervals. Specifically, for the delay $\boldsymbol{\tau}$, the prior is defined as
\vspace{-1.5mm}
\begin{equation}
p(\boldsymbol{\tau}) = \prod_{q=1}^{Q}f(\tau_q;\mu_{\uptau}, \sigma_{\uptau}), 
\label{prior_tau}
\end{equation}
\vspace{-2mm}
and we have
\begin{equation}
f(\tau_q; \mu_{\uptau}, \sigma_{\uptau}) = 
\begin{cases} 
\displaystyle \frac{\exp\left(-\frac{(\tau_q - \mu_{\uptau})^2}{2\sigma_{\uptau}^2}\right)}{\sigma_{\uptau} Z(\tau_{\mathrm{min}},\tau_{\mathrm{max}})} , & \tau_{\mathrm{min}} \leq \tau_q \leq \tau_{\mathrm{max}}, \\
0, & \mathrm{else},
\end{cases}
\end{equation}
where $Z(\tau_{\mathrm{min}},\tau_{\mathrm{max}})=\sqrt{2\pi}\big(F(\tau_{\mathrm{max}})-F(\tau_{\mathrm{min}})\big)$, and $F(\cdot)$ is the cumulative distribution function of the standard normal distribution. This formulation ensures each component $\tau_q$ is bounded over $[\tau_{\mathrm{min}},\tau_{\mathrm{max}}]$. The Doppler shift and angle parameters follow the same form as (\ref{prior_tau}) within their respective feasible detection intervals.
All unknown parameters are concatenated into a single vector $\boldsymbol{\eta} = [\boldsymbol{\tau},\boldsymbol{f}_{\mathrm{D}},\boldsymbol{\theta},\mathbf{b},\boldsymbol{\rho},\xi]^\mathrm{T}$, and the corresponding posterior distribution is given by
\begin{equation}
p(\boldsymbol{\eta}|\tilde{\mathbf{y}}) =\frac{p(\tilde{\mathbf{y}}|\boldsymbol{\eta})p(\boldsymbol{\eta})}{p(\tilde{\mathbf{y}})}.
\label{full_posterior}
\end{equation}
Assuming statistical independence among the delay, Doppler shift, angle, reflection coefficient, and noise variance, the joint prior can be factorized as
\begin{equation}
p(\boldsymbol{\eta}) = p(\boldsymbol{\tau})p(\boldsymbol{f}_{\mathrm{D}})p(\boldsymbol{\theta})p(\mathbf{b}|\boldsymbol{\rho})p(\boldsymbol{\rho})p(\xi).
\end{equation}

While the proposed gridless SBL framework addresses the trade-off between performance and complexity, it leads to a high-dimensional and non-convex posterior over parameter vector $\boldsymbol{\eta}$. Nevertheless, the MCMC discussed later is well-suited to handle such inference problems, offering strongly robust and highly accurate joint estimation.

\begin{figure*}
\centering
\includegraphics[width=0.85\linewidth]{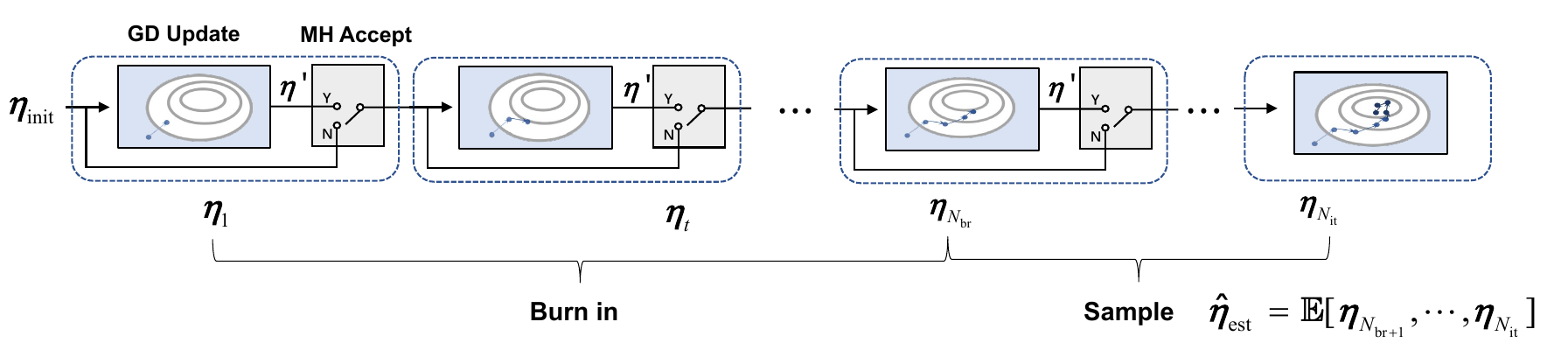}
\captionsetup{font=small}
\caption{The gradient-based MCMC sampling procedure. Each iteration consists of a proposal step and an acceptance step. The initial $N_\mathrm{br}$ samples form the burn-in phase and are discarded, while the subsequent $N_{\mathrm{sm}}$ samples are collected to compute the final estimate $\hat{\boldsymbol{\eta}}_{\mathrm{est}}$.}
\label{mcmc}
\vspace{-2mm}
\end{figure*}

\vspace{-1.5mm}
\subsection{MCMC-Based Gridless SBL Inference Algorithm}
Estimating the high-dimensional parameter vector $\boldsymbol{\eta}$ from the posterior distribution (\ref{full_posterior}) is particularly challenging due to its non-convex nature. Traditional optimization-based approaches, such as maximum a posterior estimation, tend to converge to local optima and suffer from poor scalability in such complex inference tasks.
To overcome these difficulties, we employ MCMC, which constructs an ergodic Markov chain whose stationary distribution asymptotically converges to the target distribution. 
By strategically designing the proposal mechanism that drives the Markov state transitions, MCMC generates a sequence of samples that follow the target posterior.
These samples facilitate Monte Carlo integration to approximate the estimation of parameters, while avoiding the risk of getting stuck in suboptimal regions of the parameter space.


Notably, the gradient-based MCMC has emerged as a promising solution for complex posterior inference problems \cite{jordan_faster}. These methods incorporate the gradient information of the log-posterior into the proposal mechanism, enabling more efficient state transitions in the Markov chain.  
The general form of the proposal update is given by \cite{SGLD}
\vspace{-1mm}
\begin{equation}
\boldsymbol{\eta}'
=\boldsymbol{\eta}_{t}+ \epsilon \bigtriangledown_{\boldsymbol{\eta}_t} \underbrace{\Big(\mathrm{log}p(\tilde{\mathbf{y}}|\boldsymbol{\eta}_t) +\mathrm{log}p(\boldsymbol{\eta}_t)\Big)}_{\pi(\boldsymbol{\eta}_t|\tilde{\mathbf{y}})} + \mathbf{n},
\label{GD}
\end{equation}
where $\boldsymbol{\eta}'$ is the candidate state generated with gradient information, $t$ is the index of MCMC sampling iteration, and $\boldsymbol{\eta}_t$ is the current state. $\epsilon$ is the learning rate, and $\mathbf{n} \sim \mathcal{N}(0,2\epsilon\mathbf{I})$ is the random perturbation introduced to ensure the chain explores the entire state space. $\pi(\cdot|\tilde{\mathbf{y}})$ denotes the log-unnormalized posterior, defined as the sum of the log-likelihood and log-prior. 
The logarithmic form of the posterior probability ensures that the high-dimensional integral term $p(\tilde{\mathbf{y}})$ is not explicitly computed in the gradient, while also preventing numerical overflow. 
Each proposed candidate state is evaluated using the MH acceptance rule \cite{MH_ramdom} to approximately align the Markov chain’s stationary distribution with the posterior
\begin{equation}
\alpha = \min\Bigg\{1, \frac{e^{\pi(\boldsymbol{\eta}'|\tilde{\mathbf{y}})}q(\boldsymbol{\eta}_t|\boldsymbol{\eta}')}{e^{\pi(\boldsymbol{\eta}_t|\tilde{\mathbf{y}})}q(\boldsymbol{\eta}'|\boldsymbol{\eta}_t)}\Bigg\},
\vspace{-1mm}
\end{equation}
where transition probabilities $q(\cdot|\cdot)$ induced by the proposal in (\ref{GD}) follow the complex Gaussian distribution.
This update rule depends on the acceptance probability of the candidate state $\boldsymbol{\eta}'$ is given by
\vspace{-1mm}
\begin{equation}
\boldsymbol{\eta}_{t+1} = 
\begin{cases} 
\boldsymbol{\eta}', & \text{if } \alpha \geq u \text{ (accept)}, \\
\boldsymbol{\eta}_t, & \text{otherwise (reject)},
\vspace{-1mm}
\end{cases}
\end{equation}
where $u$ is randomly sampled from uniform distribution $\mathcal{U}(0,1)$. 
Running the gradient-based MCMC for $N_\mathrm{it}$ iterations, a sample list $\boldsymbol{\eta}_1,\boldsymbol{\eta}_2,\cdots,\boldsymbol{\eta}_{N_\mathrm{it}}$ would be identified for inference. The first $N_{\mathrm{br}}$ samples constitute the burn-in phase, during which the Markov chain has not yet reached stationarity. These samples primarily serve to guide the chain to converge to the posterior. The remaining $N_{\mathrm{sm}}$ samples, where $N_{\mathrm{br}}+N_{\mathrm{sm}}=N_{\mathrm{it}}$, are drawn after the burn-in phase, indicating that the chain has reached the stationary target distribution. Posterior inference is performed using the $N_{\mathrm{sm}}$ samples with Monte Carlo integration
\vspace{-1mm}
\begin{equation}
\hat{\boldsymbol{\eta}}_{\mathrm{est}} = \frac{1}{N_{\mathrm{sm}}}\sum_{t=N_{\mathrm{br}+1}}^{N_{\mathrm{it}}}\boldsymbol{\eta}_t.
\label{average}
\vspace{-1mm}
\end{equation}

An overview of the gradient-based MCMC sampling procedure is shown in Fig. \ref{mcmc}.
By combining gradient guidance with the MH criterion, gradient-based MCMC algorithms significantly accelerate convergence to the target distribution, especially in high-dimensional and non-convex settings.
Furthermore, the logarithmic form used in the gradient computation within (\ref{GD}) and the ratio-based MH acceptance rule eliminate the need to explicitly evaluate the high-dimensional integral term $p(\tilde{\mathbf{y}})$, making them especially suitable for gridless SBL with continuous parameter spaces. 
However, despite these advantages, gradient-based MCMC methods require full-data gradient evaluations at every iteration, resluting in substantial computational latency. This constraint is particularly prohibitive for ISAC systems operating under real-time signal processing requirements. 
To address this limitation, we propose three innovative improvements to enhance sampling efficiency within the gradient-based MCMC framework.

\textsl{1) Mini-Batch Gradient}: Computing gradients is mainly dominated by the log-likelihood term in $\pi(\boldsymbol{\eta}_t|\tilde{\mathbf{y}})$, which must be evaluated  over the entire received signal dataset of size $H=MKN$ at each iteration. This cost is further amplified by the MH criterion, which makes the sampling suffering from inherent latency.
To address this challenge, we adopt a mini-batch strategy that approximates the gradient and acceptance ratio using a subset of data of size $B \ll H$. 
The update rule with the mini-batch gradient is
\vspace{-0.5mm}
\begin{equation}
\boldsymbol{\eta}'  =  \boldsymbol{\eta}_t + \epsilon_t \bigtriangledown_{\boldsymbol{\eta}_t}\bigg(\frac{H}{B}\sum_{i=1}^{B}\mathrm{log}p(y_i|\boldsymbol{\eta}_t) +\mathrm{log}p(\boldsymbol{\eta}_t)\bigg) + \mathbf{n}, 
\end{equation}
where $y_i$ is the $i$-th element of the observation vector $\tilde{\mathbf{y}}$.
The noise term is given by $\mathbf{n} \sim \mathcal{N}(\mathbf{0},2\epsilon_t\mathbf{I})$, and $\epsilon_t$ is a decaying learning rate to balance exploration and exploitation during sampling. Larger initial step sizes facilitate exploration of low-density regions to escape local optima, while reduced steps during convergence enable precise exploitation near probability maxima, enhancing both sampling efficiency and estimation accuracy.
This approach allows us to generate a proposal at a fraction of the computational cost of the full-data method.

\textsl{2) Scaling Parameter}: While mini-batch gradients reduce the computational cost of calculating gradients over the entire dataset, they introduce an approximation error that can affect the accuracy of posterior estimation. To quantify this error, we define the average log-likelihood over the full dataset as
\begin{equation}
\mu(\boldsymbol{\eta}_t) = \sum_{i=1}^{H}\mathrm{log} p(y_i|\boldsymbol{\eta}_t), 
\label{posterior}
\vspace{-2mm}
\end{equation}
and when using mini-batch gradients, we approximate the log-likelihood as
\begin{equation}
\vspace{-2mm}
\tilde{\mu}(\boldsymbol{\eta}_t) = \frac{H}{B}\sum_{i=1}^{B}\mathrm{log} p(y_i|\boldsymbol{\eta}_t).
\end{equation}
Obviously, $\tilde{\mu}(\boldsymbol{\eta})$ is the unbiased estimator of the log-likelihood $\mu(\boldsymbol{\eta})$, such that $\mathbb{E}[\tilde{\mu}(\boldsymbol{\eta}_t)]=\mu(\boldsymbol{\eta}_t)$. 
However, the MH acceptance criterion does not operate on the log-likelihood directly, but on the likelihood itself, which is computed via an exponential function. Due to the non-linearity of this function, the unbiased property is lost. Specifically, Jensen's inequality implies that
	\vspace{-3mm}
\begin{equation}
	\vspace{-1mm}
\mathbb{E}[e^{\tilde{\mu}(\boldsymbol{\eta})}]\neq e^{\mathbb{E}[\tilde{\mu}(\boldsymbol{\eta})]}= e^{\mu(\boldsymbol{\eta}_t)}.
\end{equation}
This inequality demonstrates that a naive MH ratio using a mini-batch likelihood estimate is fundamentally biased, causing the sampler's stationary distribution to deviate significantly from the true target posterior in (\ref{full_posterior}). 
Inspired by \cite{temperature}, we introduce a scaling parameter $\gamma$ to mitigate this bias, and the mini-batch gradient is modified as
\vspace{-1mm}
\begin{equation}
\boldsymbol{\eta}' \! = \! \boldsymbol{\eta}_t + \epsilon_t \bigtriangledown_{\boldsymbol{\eta}_t}\underbrace{\bigg(\frac{H^{\gamma}}{B}\sum_{i=1}^{B}\mathrm{log}p(y_i|\boldsymbol{\eta}_t) +\mathrm{log}p(\boldsymbol{\eta}_t)\bigg)}_{\check{\pi}(\boldsymbol{\eta}_t|\tilde{\mathbf{y}})}+ \mathbf{n},
\label{mbgradient}
\end{equation}
where $\gamma \triangleq \upsilon\mathrm{log}B/\mathrm{log}H $, and $ \upsilon \in (0,1)$ is a tunable hyperparameter. Consequently, the MH acceptance criterion can be computed efficiently with mini-batch likelihood as 
\begin{equation}
\alpha = \min\Bigg\{1, \frac{e^{\check{\pi}(\boldsymbol{\eta}'|\tilde{\mathbf{y}})}}{e^{\check{\pi}(\boldsymbol{\eta}_t|\tilde{\mathbf{y}})}}\Bigg\},
\label{ratio}
\vspace{-1mm}
\end{equation}
where transition probabilities are omitted for simplicity \cite{zhou_minibatchmcmc}.
The method's ingenuity lies in converting the uncontrolled bias of mini-batching a predictable feature. A naive MH ratio with mini-batch likelihood leads to an intractable stationary distribution dependent on the estimator's variance. In contrast, the correction of $\gamma$ reshapes this uncontrolled bias into a predictable tempering effect, guaranteeing that the sampler converges to an analytically tractable tempered version of (\ref{full_posterior})
\begin{equation}
	\vspace{-1.5mm}
p_{\gamma}(\boldsymbol{\eta}_t|\tilde{\mathbf{y}}) \propto  p(\tilde{\mathbf{y}}|\boldsymbol{\eta}_t)^{H^{\gamma-1}}p(\boldsymbol{\eta}_t).
\end{equation} 
This adaptive scaling mechanism offers significant advantages. When the mini-batch size $B$ is large, $\gamma$ increases, causing the exponent $H^{\gamma-1}$ to approach 1 and ensuring the stationary distribution faithfully approximates the true posterior. Conversely, for a smaller $B$, the exponent is reduced, flattening the distribution to enhance exploration. Crucially, since tempering preserves the locations of the posterior modes, this enhanced exploration does not compromise the accuracy of the final point estimate, which is averaged from converged samples.

\begin{algorithm}[t]
\caption{MCMC-Based Gridless SBL}
\label{al1}
\begin{algorithmic}[1]
\REQUIRE Randomly initial parameters $\boldsymbol{\eta}_{0}$, learning rate $\epsilon_{0}$, mini-batch size $B$, sample set $\boldsymbol{\Psi}=\emptyset$, burn-in count $N_{\mathrm{br}}$, sample number $N_{\mathrm{sm}}$, and the total iteration number $N_{\mathrm{it}}$.
\ENSURE Estimated parameters $\hat{\boldsymbol{\eta}}_{\mathrm{est}}=[\hat{\boldsymbol{\tau}},\hat{\boldsymbol{f}_\mathrm{D}},\hat{\boldsymbol{\theta}},\hat{\mathbf{b}},\hat{\boldsymbol{\rho}},\hat{\xi}]^{\mathrm{T}}$.

\FOR{step $t=1$ to $N_{\mathrm{it}}$}
\STATE Randomly sample a mini-batch of received signal;
\STATE Compute the gradient $\mathbf{g}_t$ over the mini-batch;
\STATE Construct the candidate parameters vector $\boldsymbol{\eta}'$ with Adam optimizer via (\ref{eq14});  
\IF{$\boldsymbol{\eta}'$ is out of the range}
\STATE $\boldsymbol{\eta}_{t+1}=\boldsymbol{\eta}_{t}$;
\STATE Continue;
\ELSE
\STATE Compute the acceptance probability $\alpha$ via (\ref{ratio});
\STATE Generate a random number $u$ from $\mathcal{U}(0,1)$;
\IF{$\alpha \geq u $}
\STATE $\boldsymbol{\eta}_{t+1}=\boldsymbol{\eta}'$;
\ELSE
\STATE $\boldsymbol{\eta}_{t+1}=\boldsymbol{\eta}_{t}$;
\ENDIF
\IF{$t \geq N_{\mathrm{br}}$}
\STATE Add $\boldsymbol{\eta}_{t+1}$ to sample set $\boldsymbol{\Psi}$;
\ENDIF
\ENDIF
\ENDFOR
\STATE Compute the average value $\hat{\boldsymbol{\eta}}_{\mathrm{est}}$ over $\boldsymbol{\Psi}$ via (\ref{average}).
\end{algorithmic}
\end{algorithm}

\textsl{3) Adam-Based Sampler}: 
To enhance sampling efficiency in high-dimensional, anisotropic posterior landscapes, particularly with noisy mini-batch gradients, we employ a sophisticated proposal mechanism inspired by the Adam optimizer \cite{Kingma2014AdamAM}. This approach leverages momentum to smooth the sampling trajectory and adaptive preconditioning to stabilize updates.
Specifically, we maintain exponentially decaying averages of the first ($\mathbf{w}_t$, the momentum term) and the second ($\mathbf{v}_t$, the uncentered variance) moments of the tempered mini-batch gradients $\mathbf{g}_t=\bigtriangledown_{\boldsymbol{\eta}_t} \check{\pi}(\boldsymbol{\eta}_t|\tilde{\mathbf{y}})$ with
\begin{equation}
\vspace{-1mm}
\begin{aligned}
&\mathbf{w}_t=\beta_1\mathbf{w}_{t-1}+(1-\beta_1)\mathbf{g}_t,\\
&\mathbf{v}_t=\beta_2\mathbf{v}_{t-1}+(1-\beta_2)\mathbf{g}_t \odot \mathbf{g}_t ,
\label{eq13}
\end{aligned}
\end{equation}
where $\beta_1, \beta_2 \in [0,1)$ are attenuation coefficients. The bias-corrected moments are calculated as
\begin{equation}
\hat{\mathbf{w}}_{t} = \frac{\mathbf{w}_{t}}{1-\beta_{1}^{t}}, \quad \hat{\mathbf{v}}_{t} = \frac{\mathbf{v}_{t}}{1-\beta_{2}^{t}},
\end{equation}
which are combined to form the Adam-based proposal as
\begin{equation}
\boldsymbol{\eta}' =\boldsymbol{\eta}_{t}+\epsilon_t \hat{\mathbf{w}}_t\oslash (\sqrt{\hat{\mathbf{v}}_t}+\varepsilon) + \mathbf{n},
\label{eq14}
\end{equation}
where $\varepsilon$ is a small constant for numerical stability. Crucially, the injected noise $\mathbf{n}$ is also preconditioned by the geometry learned by Adam. Instead of isotropic noise, it is drawn from a zero-mean Gaussian with an adaptive diagonal covariance
\begin{equation}
\mathbf{n} \sim \mathcal{CN} \Big(\mathbf{0}, \frac{2\epsilon_t}{ \sqrt{\hat{\mathbf{v}}_t} + \varepsilon}\Big).
\end{equation}
This adaptive preconditioning of both the gradient and the random noise makes the sampler highly efficient. It encourages larger exploratory steps in flat regions of the parameter space and smaller, more cautious steps in steep regions, leading to faster convergence and more stable sampling when using noisy mini-batch gradients.

The complete MCMC sampling algorithm is summarized in Algorithm \ref{al1}. Notably, the majority of parameters in $\boldsymbol{\eta}_t$ are bounded. If any parameters in $\boldsymbol{\eta}_{t+1}$ exceed their specified range during sampling, the entire $\boldsymbol{\eta}_{t+1}$ will be rejected outright.
\begin{figure*}
\centering
\includegraphics[width=0.85\linewidth]{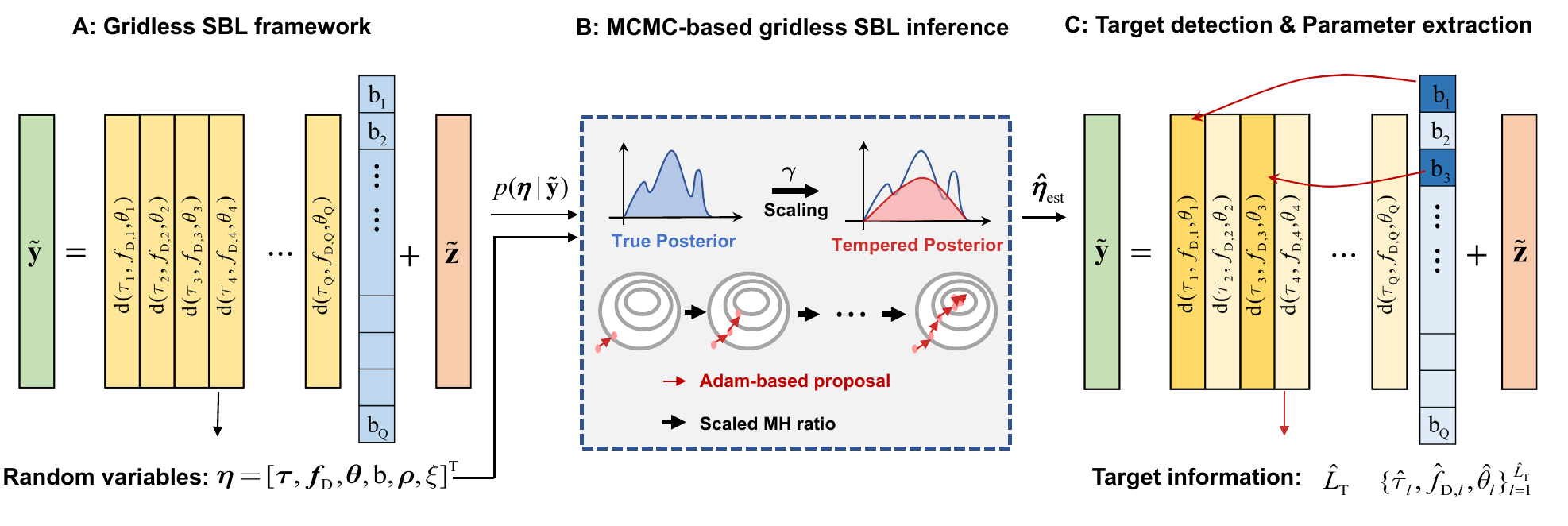}
\captionsetup{font={small}}
\caption{Workflow of the proposed MCMC-based gridless SBL algorithm. A: The gridless SBL model is formulated, treating target parameters $(\boldsymbol{\tau}, \boldsymbol{f}_{\mathrm{D}},\boldsymbol{\theta})$ as continuous random variables. B: An efficient MCMC sampler is employed to perform Bayesian inference, generating $\hat{\boldsymbol{\eta}}_{\mathrm{est}}$ from the tempered posterior $p_{\gamma}(\boldsymbol{\eta}|\tilde{\mathbf{y}})$. C: Target detection is performed by identifying the dominant coefficients in $\hat{\mathbf{b}}$, and the corresponding continuous parameters are extracted to obtain the final estimates.}
\label{sbl_workflow}
\vspace{-3mm}
\end{figure*}

\vspace{-4mm}
\subsection{Target Detection and Parameter Extraction}
Upon completion of the MCMC sampling in Algorithm \ref{al1}, we obtain the parameter mean estimation $\hat{\boldsymbol{\eta}}_{\mathrm{est}}$, which contain all the necessary information for both target detection and corresponding parameters extraction.
A key advantage of our MCMC-based gridless SBL framework is its inherent ability to perform automatic model order selection, without the need for a pre-specified target count required by many conventional CS methods.
This is realized through the hierarchical prior placed on the reflection coefficients $\mathbf{b}$. During MCMC inference, the posterior distributions of coefficients corresponding to spurious or non-existent targets naturally collapse towards zero. 
Consequently, the task of target detection reduces to identifying the significant non-zero elements within the estimated coefficient vector $\hat{\mathbf{b}}$.
We formalize this using an energy-based thresholding criterion.
The number of detected targets $\hat{L}_{\mathrm{T}}$ is determined by finding the smallest positive integer $h$ that satisfies a cumulative energy constraint
\vspace{-1mm}
\begin{equation}
\begin{aligned}
\hat{L}_\mathrm{T} =& \min_{h} \quad h \\
&\text{s.t.} \quad \frac{\sum_{i=1}^{h} |\hat{b}_{(i)}|^2}{\sum_{q=1}^{Q} |\hat{b}_q|^2} \geq \sigma_{\mathrm{thr}}, \\
&\quad \quad h \in \mathbb{Z}^+
\end{aligned}
\vspace{-1mm}
\end{equation}
where $\{|\hat{b}_{(i)}|\}_{i=1}^Q$ is the sequence of elements from $\hat{\boldsymbol{b}}$ sorted by their magnitudes in descending order, such that $|\hat{b}_{(1)}| \geq|\hat{b}_{(2)}| \geq \cdots \geq |\hat{b}_{(Q)}|$. The denominator $\sum_{q=1}^{Q} |\hat{b}_q|^2$ represents the total energy of the estimated $\hat{\boldsymbol{b}}$.
The energy threshold $\sigma_{\mathrm{thr}}$ is set to a value that provides a robust empirical balance between capturing the energy of true targets and rejecting spurious, low-energy estimates arising from noise.

Once $\hat{L}_{\mathrm{T}}$ is determined, the indices of these $\hat{L}_{\mathrm{T}}$ dominant coefficients identify the active targets. 
Instead of a post-hoc search through a large, discretized dictionary as in conventional CS, our framework jointly estimates each reflection coefficient $\hat{{b}}_q$ and its corresponding continuous-valued parameter set ($\hat{\tau}_q,\hat{f}_{\mathrm{D},q},\hat{\theta}_q$). This estimation is performed over a compact model space of dimension $Q$, which is chosen to only slightly exceed the expected number of targets. Therefore, parameter extraction becomes a simple direct lookup.
The physical parameters for each of the $\hat{L}_{\mathrm{T}}$ detected targets $\{\hat{\tau}_l,\hat{f}_{\mathrm{D},l},\hat{\theta}_l\}_{l=1}^{\hat{L}_{\mathrm{T}}}$ are then extracted from the corresponding entries in the estimated parameter vectors $\hat{\boldsymbol{\tau}},\hat{\boldsymbol{f}}_{\mathrm{D}},\hat{\boldsymbol{\theta}}$. 
The overall framework of MCMC-based gridless SBL algorithm is shown in Fig. \ref{sbl_workflow}.

\subsection{Bayesian Cramér–Rao Bound}
Having established the parameter estimation procedure, it is crucial to benchmark its performance against a theoretical limit. We derive the BCRB, which is particularly suitable for our SBL framework as it incorporates prior information to provide a tighter bound than the classical CRB for Bayesian estimators \cite{bcrb_conf}. 
Our analysis focuses on the key physical parameters, reorganized into a single real-valued vector $\boldsymbol{\zeta}=[\mathbf{b}_{\mathrm{R}},\mathbf{b}_{\mathrm{I}},\boldsymbol{\tau},\boldsymbol{f}_{\mathrm{D}},\boldsymbol{\theta},\xi]^{\mathrm{T}}$. Note that the hyperparameter $\boldsymbol{\rho}$, being the latent variable for inducing sparsity, is excluded from this BCRB calculation.
The BCRB states that the MSE matrix for any Bayesian estimator $\hat{\zeta}$ is lower-bounded by the inverse of the Bayesian Fisher information matrix (BFIM) $\boldsymbol{\mathcal{I}}_{\mathrm{B}}$
\begin{equation}
\mathbb{E}\left[(\boldsymbol{\zeta} - \hat{\boldsymbol{\zeta}})(\boldsymbol{\zeta} - \hat{\boldsymbol{\zeta}})^{\mathrm{T}}\right] \succeq \boldsymbol{\mathcal{I}}_{\mathrm{B}}^{-1}.
\label{eq:defination of BCRB}
\end{equation}
Consequently, the BCRB is given by the trace of this matrix 
\begin{equation}
\mathrm{BCRB}(\boldsymbol{\zeta}) = \mathrm{Tr}\left(\boldsymbol{\mathcal{I}}_\mathrm{B}^{-1}\right).
\label{BCRB_def}
\end{equation}
The BFIM is composed of the classical FIM $\boldsymbol{\mathcal{I}}_{\mathrm{C}}$ purely from the observation data, and prior information matrix $\boldsymbol{\mathcal{I}}_{\mathrm{P}}$
\begin{equation}
\vspace{-2mm}
\boldsymbol{\mathcal{I}}_{\mathrm{B}} = \mathbb{E}_{\boldsymbol{\zeta}}[\boldsymbol{\mathcal{I}}_{\mathrm{C}}] +\boldsymbol{\mathcal{I}}_{\mathrm{P}},
\label{BFIM_com}
\end{equation}
where $\mathbb{E}_{\boldsymbol{\zeta}}[\boldsymbol{\mathcal{I}}_{\mathrm{C}}]$  is typically approximated via Monte Carlo integration over the prior distributions (see in (\ref{MC_BFIM})).
Analyzing these two components reveals the physical origins of estimation accuracy and the theoretical basis of our SBL approach. 

As detailed in Appendix A, the classical FIM $\boldsymbol{\mathcal{I}}_{\mathrm{C}}$ exhibits a block-diagonal structure (derived in (\ref{diag_Ic})), decoupling the physical parameters from the inverse noise variance $\xi$. The FIM block for the physical parameters $\boldsymbol{\mathcal{I}}_{\mathrm{C}}^s$ is
\vspace{-2mm}
\begin{equation}
\boldsymbol{\mathcal{I}}_{\mathrm{C}}^{s} = \sum_{r=1}^{M \times K} 2\xi \cdot \mathrm{Re}\left\{ \mathbf{p}_r^* \left( \mathbf{v}_r^{\mathrm{H}} (\mathbf{x}_r^* \mathbf{x}_r^{\mathrm{T}} \otimes \mathbf{I}) \mathbf{v}_r \right) \mathbf{p}_r^{\mathrm{T}} \right\}, 
\vspace{-1mm}
\end{equation}
where $ \mathbf{v}_r$ contains the parameters and $\mathbf{p}_r$ is a modulation vector. This reveals that the best achievable precision is fundamentally limited by the system's physical configuration
\begin{equation}
[\boldsymbol{\mathcal{I}}_{\mathrm{C}}^s]_{\tau_l.\tau_l}\! \propto B^2, \quad \! \!
[\boldsymbol{\mathcal{I}}_{\mathrm{C}}^s]_{f_{\mathrm{D},l},f_{\mathrm{D},l}} \! \propto T_{\mathrm{total}}^2,\quad \! \!
[\boldsymbol{\mathcal{I}}_{\mathrm{C}}^s]_{\theta_l,\theta_l}\! \propto D^2,
\end{equation}
where $B=M\Delta f$, $T_{\mathrm{total}}=K T_{\mathrm{s}}$, and $D=\lambda(N-1)/2$ are the signal bandwidth, total duration, and array aperture, respectively.
However, this data-dependent theoretical bound becomes problematic in challenging scenarios. In cases such as low SNR or super-resolution, the information provided by data is either scarce or ambiguous. This causes the classical FIM to become ill-conditioned, rendering the corresponding CRB (derived from $\mathrm{Tr}(\boldsymbol{\mathcal{I}}_{\mathrm{C}}^{-1})$) to lose practical relevance due to numerical instability. This indicates the failure of non-Bayesian estimators and necessitates the use of prior information.

Our SBL model resolves this by introducing the prior information matrix $\boldsymbol{\mathcal{I}}_{\mathrm{P}}$. The priors on the physical parameters result in a positive-definite diagonal matrix with entries $[\boldsymbol{\mathcal{I}}_{\mathrm{P}}]_{*,*}=1/\sigma^2_{*}$, for $*\in {\tau_l,f_{\mathrm{D},l},\theta_l}$.
This matrix regularizes the BFIM to ensure it remains well-conditioned and invertible.
Consequently, the BCRB provides a meaningful performance bound determined by the prior knowledge, even when the CRB becomes impractical. This regularization is the theoretical cornerstone of our algorithm's ability to achieve robust, super-resolution performance. 
Our MCMC sampler then efficiently finds a near-optimal solution to track this bound.

\vspace{-3mm}
\subsection{Complexity Analysis}
In this subsection, we analyze the complexity of the proposed MCMC-based SBL algorithm from two perspectives: computational complexity and storage complexity, and compare it with conventional grid-based CS methods.

\textsl{1) Computation}: The computational complexity per iteration is dominated by three components: gradient calculation, parameter updates via the Adam-based optimizer, and the MH ratio. Let the total number of parameters be $F = 6Q + 1$, where the $Q$ complex coefficients $\mathbf{b}$ are decoupled into 2Q real and imaginary components.
In each iteration, both the gradient and the MH acceptance ratio are approximated using a mini-batch of $B$ data points, leading to a computational complexity of $\mathcal{O}(FB)$. The Adam optimizer needs to maintain the momentum term and scaling term, and the complexity of updating these terms linearly related to the parameter dimension is $\mathcal{O}(F)$.  
Therefore, the overall per-iteration computational complexity is $\mathcal{O}(FB)$, and the total complexity over $N_{\mathrm{it}}$ iterations is $\mathcal{O}(FN_{\mathrm{it}}B)$. 
Compared to traditional full-data MCMC, whose complexity is $\mathcal{O}(F N_{\mathrm{it}} M K N)$, the proposed mini-batch strategy significantly reduces the computational load, especially when $B \ll MKN$.
Furthermore, traditional CS methods \cite{omp} require the construction of an overcomplete dictionary and typically involve matrix inversion \cite{nomp}, leading to the computational costs as high as $\mathcal{O}((MKN)^3)$ in the worst case, which is significantly higher than our algorithm.

\textsl{2) Storage}:  The storage complexity of our algorithm is primarily determined by two components: the memory required to store the dictionary $\mathbf{D}$ and the storage for the final samples. Our algorithm's main storage cost comes from retaining the $N_{\mathrm{sm}}$ collected samples after the burn-in phase, which amounts to $\mathcal{O}(FN_{\mathrm{sm}})$.
Unlike grid-based CS methods, our approach avoids pre-computing and storing a massive, fixed dictionary. As a result, the dictionary-related memory requirement is reduced to $\mathcal{O}(QMKN)$. Therefore, the total storage complexity of our method is $\mathcal{O}(QMKN + FN_{\mathrm{sm}})$.
Traditional grid-based CS methods \cite{nomp} require storing an overcomplete dictionary of size $GMNK$, where $G$ is the total number of grid points. If the delay, Doppler, and angle dimensions are each discretized into $P$ levels with $G = P^3$, leading to a storage complexity of $\mathcal{O}(P^3MKN)$. As the desired precision increases, the storage demand grows cubically, quickly becoming intractable.

\vspace{-1mm}

\section{Numerical Results}
In this section, we present simulation results to validate the performance of the proposed MCMC-based gridless SBL algorithm.
We consider a ground-based ISAC system surveilling a mixed air-ground environment, where the simulated multi-target scene contains both low-altitude targets (e.g., UAVs) and ground targets (e.g., high-speed vehicles). 
Parameters of each target are generated following the uniform distribution of $r \in [50, 600]$ m for distance, $v \in [-30,-5] \cup [5,30]$ m/s for velocity, and $\theta \in [-80, 80]^{\circ}$ for angle. 
The amplitude of the complex reflection coefficient is simulated according to the radar equation, $b_l \propto \sqrt{\lambda^2\sigma_\mathrm{RCS}/r^4}$, where we assume a larger RCS $\sigma_{\mathrm{RCS}}$  for ground vehicles than for smaller UAVs, and the phase is uniformly random. The maximum number of $L_{\mathrm{T}}$ is set to 5 to simulate a moderately dense environment. 
We assume a quasi-static channel where target parameters remain constant for a short duration. This period is assumed to last until a target moves approximately 3 cm at 30 m/s, corresponding to a coherent processing interval of 1 ms, or about 100 OFDM symbols. 
To leverage the scene sparsity while accommodating potential targets, we set the sparse variant dimension as $Q=10$ in our MCMC-based gridless SBL algorithm, which is significantly larger than the target number $L_{\mathrm{T}}$. The relationship between the learning rate and the sampling steps is $\epsilon_t = \epsilon_0/(1+t^{0.005})$. Other system settings and algorithm parameters are summarized in Table \ref{table1} and Table \ref{table2}, respectively.

\vspace{-1mm}
\begin{table}[t!]
\centering
\captionsetup{font={small}}
\caption{System and Signal Parameters for LAWNs}
\begin{tabular}{m{3.5cm}<{\centering}m{3cm}<{\centering}}
	\toprule
	\textbf{System Parameter} & \textbf{Value} \\ 
	\midrule
	Number of antennas $(N)$ & 8 \\ 
	Antenna spacing $(d)$ & $\lambda/2$ \\
	Carrier frequency $(f_\mathrm{c})$ & 30 GHz \\ 
	Number of subcarriers $(M)$ & 128 \\ 
	Subcarrier bandwidth $(\Delta f)$ & 120 kHz \\ 
	CP length $(T_{\mathrm{cp}})$ & $\frac{1}{4} T_\mathrm{d}$ \\
	OFDM symbols $(K)$ & 14 \\ 
	\bottomrule
\end{tabular}
\label{table1}
\end{table}
\vspace{-1mm}
\begin{table}[t!]
\centering
\captionsetup{font={small}}
\caption{MCMC Algorithm Parameters}
	\begin{tabular}{m{3.5cm}<{\centering}m{3cm}<{\centering}}
		\toprule
		\textbf{Algorithm Parameter} & \textbf{Value} \\ 
		\midrule
		Mini-batch size $(B)$ & 128 \\ 
		Initial learning rate $(\epsilon_0)$ & 0.15\\
		Adam coefficients $(\beta_1$, $\beta_2)$ & 0.9, 0.9999 \\ 
		Adam stability term $(\varepsilon)$ & $10^{-8}$ \\ 
		Burn-in steps $(N_{\mathrm{br}})$ & $4 \times 10^4$ \\ 
		Sample steps $(N_{\mathrm{sm}})$ & 5000 \\ 
		\bottomrule
	\end{tabular}
	\label{table2}
\end{table}

\vspace{-0.5mm}
\subsection{Dynamic Target Detection}
\begin{figure}[t]
	\centering
	\vspace{-2mm}
	\includegraphics[width=0.85\linewidth]{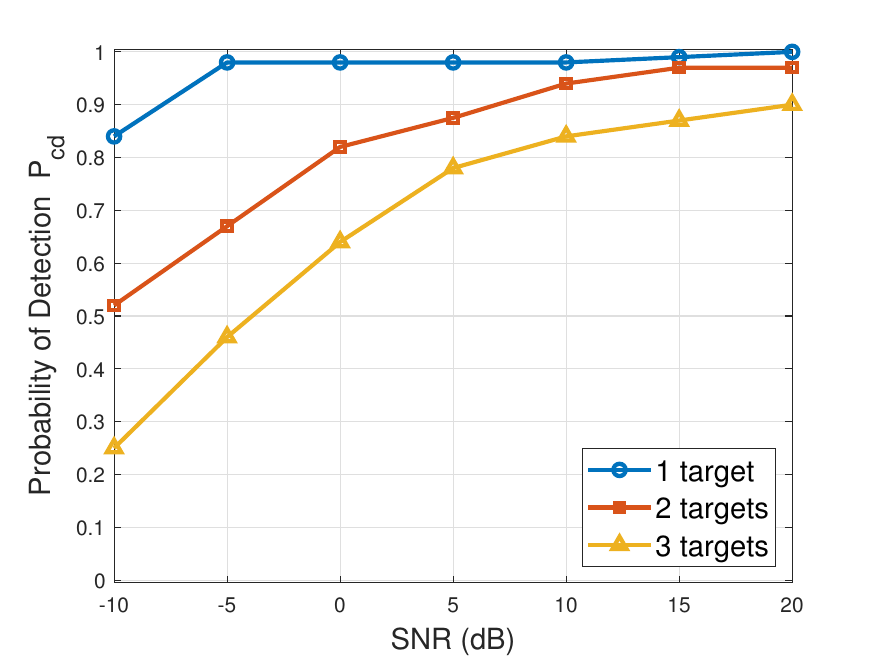}
	\captionsetup{font={small}}
	\caption{Probability of correct detection $P_{\mathrm{cd}}$ versus SNR for different numbers of dynamic targets.}
	\label{Pcd}
	\vspace{-3mm}
\end{figure}

We evaluate the target detection performance by measuring the probability of correct detection $P_{\mathrm{cd}}$, defined as the event where all true targets are successfully identified in multi-target scenarios. The detection threshold $\sigma_{\mathrm{thr}}$ is chosen empirically as 0.9.
Fig. \ref{Pcd} depicts the performance curve of the correct detection probability $P_{\mathrm{cd}}$ versus SNR for different numbers of targets, with results averaged over 200 Monte Carlo trials.

The results highlight two key strengths of our proposed algorithm: high sensitivity and robustness in multi-target environments.
High sensitivity is demonstrated by the algorithm's ability to achieve near-optimal detection
($P_{\mathrm{cd}} \to 95\%$) at high SNRs across all tested scenarios. In the single-target case, the algorithm shows excellent performance even at a low SNR of -10 dB, achieving a remarkable detection rate of approximately $85\%$. This showcases its effectiveness in noise-limited conditions.
The algorithm's robustness is evident as the number of targets increases. Although the detection rate predictably decreases with the introduction of more targets due to inter-target interference, our algorithm maintains strong performance. For instance, with three targets, it achieves a detection rate of approximately $65\%$ at an SNR of 0 dB, and surpasses $80\%$ once the SNR reaches 10 dB. This confirms its suitability for practical, dense LAWNs scenarios.

\vspace{-1mm}
\begin{figure*}
	\centering
	\vspace{-2mm}
	\begin{subfigure}[t]{0.32\textwidth}
		\centering
		\includegraphics[width=\textwidth]{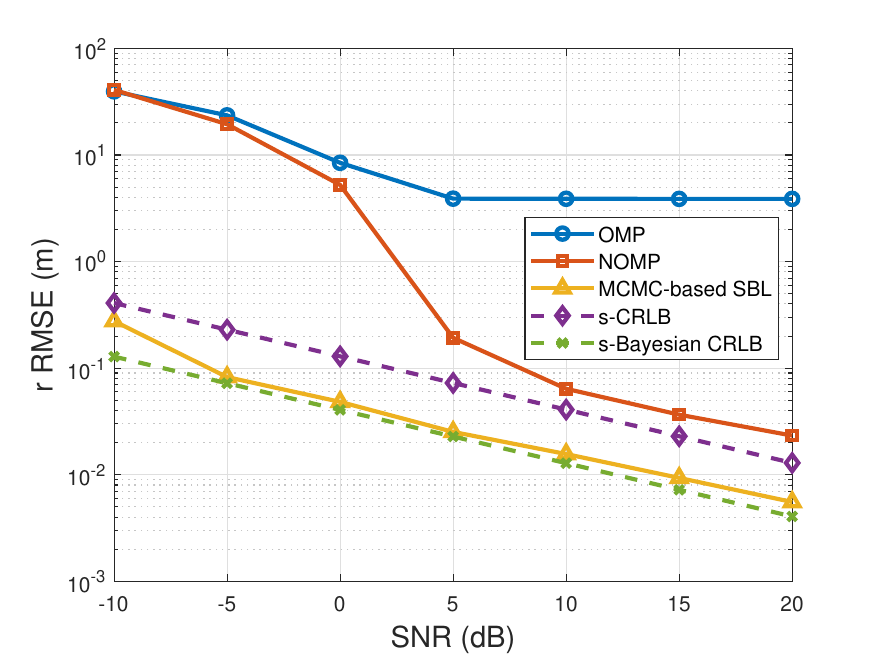}
		\captionsetup{font={small}}
		\caption{The RMSE of range estimation}
		\label{t1_range}
	\end{subfigure}
	\hfill
	\begin{subfigure}[t]{0.32\textwidth}
		\centering
		\includegraphics[width=\textwidth]{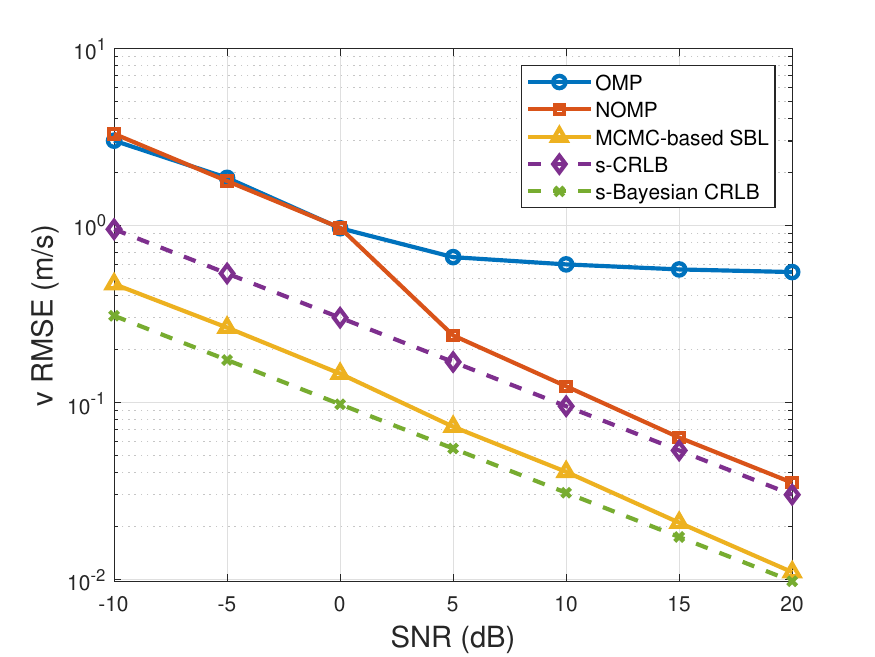} 
		\captionsetup{font={small}}
		\caption{The RMSE of velocity estimation}
		\label{t1_velocity}
	\end{subfigure}
	\hfill
	\begin{subfigure}[t]{0.32\textwidth}
		\centering
		\includegraphics[width=\textwidth]{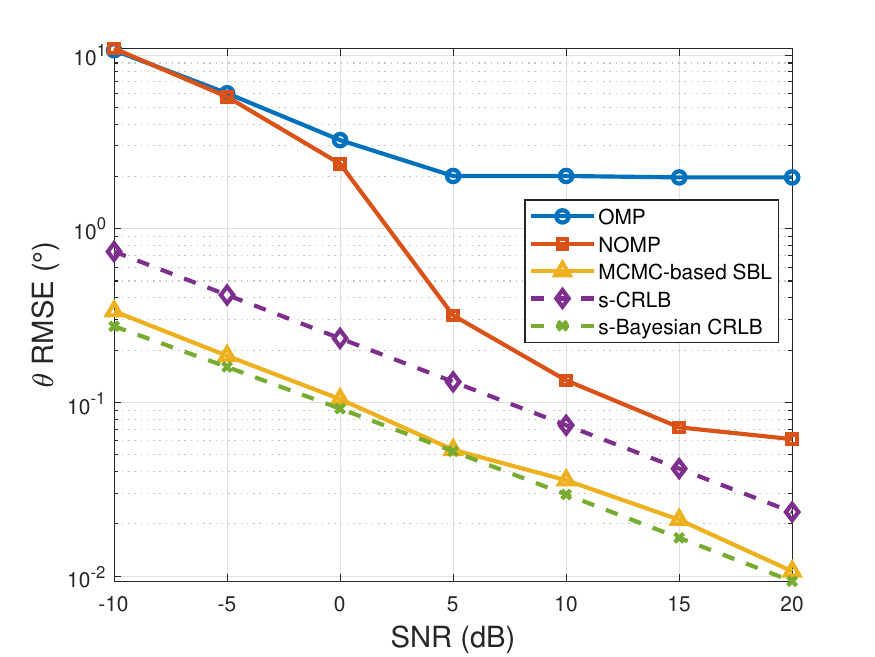}
		\captionsetup{font={small}}
		\caption{The RMSE of angle estimation}
		\label{t1_angle}
	\end{subfigure}
	\captionsetup{font={small}}
	\caption{Single-target estimation RMSE versus SNR.}
	\label{baseline_t1}
	\vspace{-4mm}
\end{figure*}
\vspace{-1mm}
\subsection{Dynamic Target Parameter Estimation}
\begin{figure*}
	\centering
	\vspace{-1mm}
	\begin{subfigure}[t]{0.32\textwidth}
		\centering
		\includegraphics[width=\textwidth]{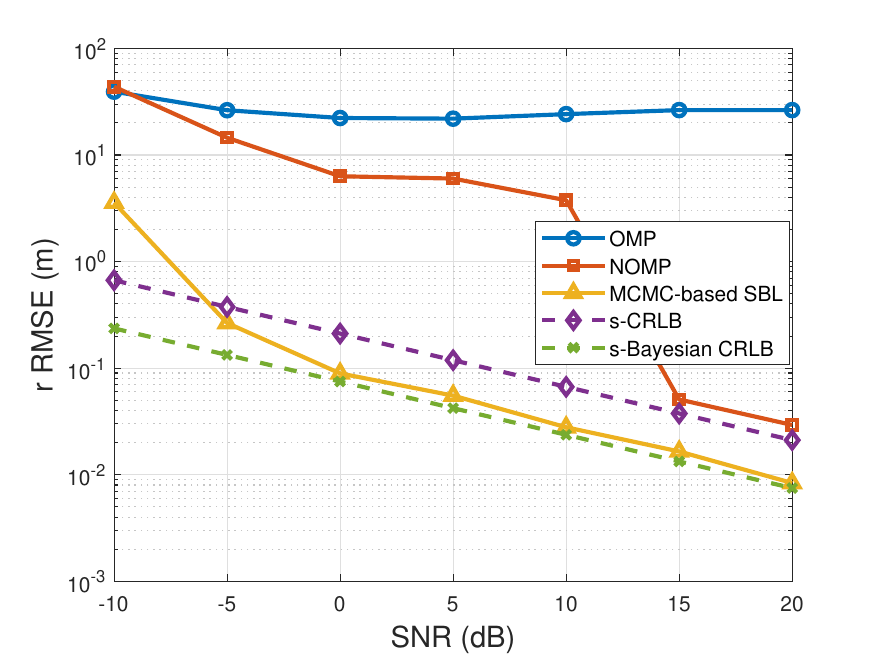}
		\captionsetup{font={small}}
		\caption{The RMSE of range estimation}
		\label{t3_range}
	\end{subfigure}
	\hfill
	\begin{subfigure}[t]{0.32\textwidth}
		\centering
		\includegraphics[width=\textwidth]{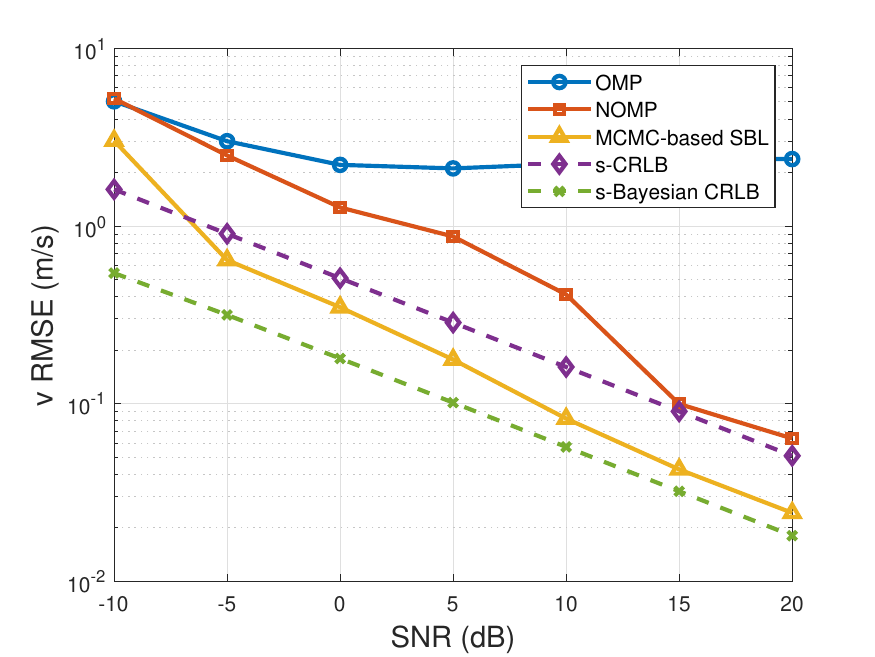} 
		\captionsetup{font={small}}
		\caption{The RMSE of velocity estimation}
		\label{t3_velocity}
	\end{subfigure}
	\hfill
	\begin{subfigure}[t]{0.32\textwidth}
		\centering
		\includegraphics[width=\textwidth]{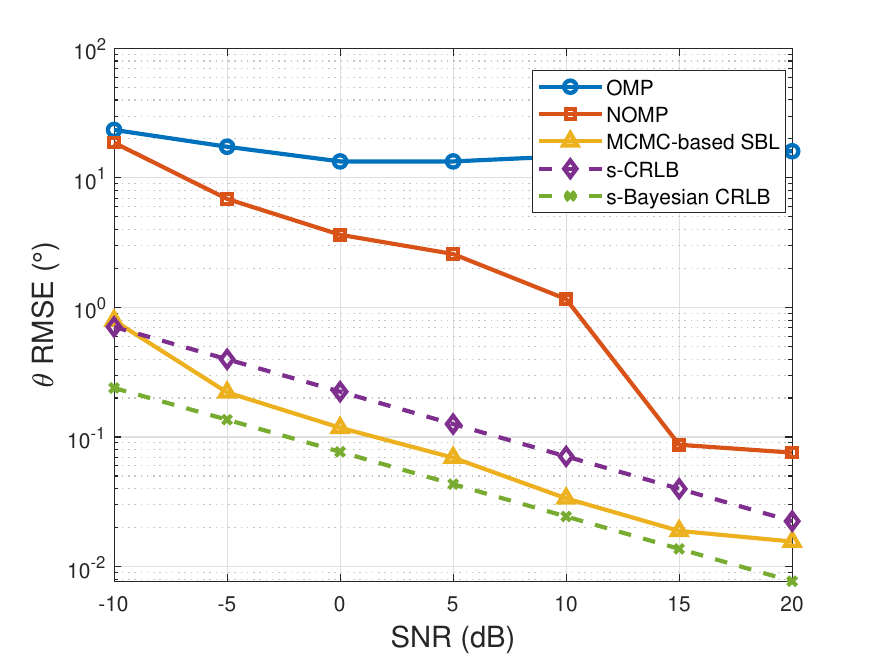}
		\captionsetup{font={small}}
		\caption{The RMSE of angle estimation}
		\label{t3_angle}
	\end{subfigure}
	\captionsetup{font={small}}
	\caption{Multiple-targets estimation RMSE versus SNR.}
	\label{baseline_t3}
	\vspace{-2mm}
\end{figure*}

Fig. \ref{baseline_t1} and Fig. \ref{baseline_t3} show the root mean square error (RMSE) for range, velocity, and angle estimation versus SNR in both single-target and multi-target scenarios. We compare the performance of the proposed MCMC-based gridless SBL algorithm against traditional grid-based CS algorithms (on-grid OMP \cite{omp} and off-grid NOMP \cite{nomp}) and present the square root bounds, s-BCRB and s-CRB, for both algorithm types, respectively.

In the single-target case (Fig. \ref{baseline_t1}), our proposed algorithm demonstrates a commanding performance advantage across the entire SNR spectrum.
In low SNR scenarios, (e.g., -10 dB), it achieves a profound improvement, outperforming the NOMP baseline by approximately 20 dB in range, 14 dB in velocity, and 20 dB in angle estimation. 
This robustness stems from the Bayesian framework's ability to precisely model noise variance via (\ref{noise_variance}), a feature lacking in noise-sensitive matching pursuit methods like OMP and NOMP. 
As SNR increases, the algorithm's superiority becomes even more evident by its ability to achieve ultra-high estimation precision, reaching 0.005 m in range, 0.01 m/s in velocity, and 0.01° in angle at 20 dB.
This persistent performance gap underscores fundamental limitations in grid-dependent approaches: OMP suffers inherent grid mismatch errors of the dictionary, and although NOMP refines grid error through Newton's method, it still suffers from exponential growth in computational complexity and struggles to achieve optimal search in continuous parameter spaces.
Most critically, our algorithm's performance curve tightly tracks the theoretical BCRB within a mere 0.5 dB, confirming its near-optimal estimation capability. This tight alignment contrasts sharply with baseline methods that reference the looser classical CRB, which underestimates achievable precision in Bayesian estimation frameworks. 

In the more challenging three-target scenario (Fig. \ref{baseline_t3}), our algorithm showcases its exceptional robustness and scalability. While conventional methods struggle with inter-target interference, our method maintains a significant lead, exhibiting a 15 dB advantage over baselines at -10 dB.
At high SNR (20 dB), it sustains ultra-high precision, achieving accuracies of 0.07 m, 0.024 m/s, and $0.015^{\circ}$ and outperforming NOMP by 14 dB, 14 dB, and 8 dB, respectively. 
Crucially, the algorithm achieves these gains without prior knowledge of target count, automatically adapting to variations in target number through its continuous parameter space optimization. 
This seamless scalability addresses a fundamental limitation of CS methods that require explicit target number initialization.
The uniform accuracy distribution across all three targets highlights the framework's robustness against target interference. This stability results from the joint multi-dimensional optimization that simultaneously resolves range-velocity-angle parameters without dimensional bias. These capabilities establish a new paradigm for ISAC systems operating in target-dense environments where conventional methods suffer from cascading estimation errors.

\begin{figure*}
	\centering
	\vspace{-2mm}
	\begin{subfigure}[t]{0.32\textwidth}
		\centering
		\includegraphics[width=\textwidth]{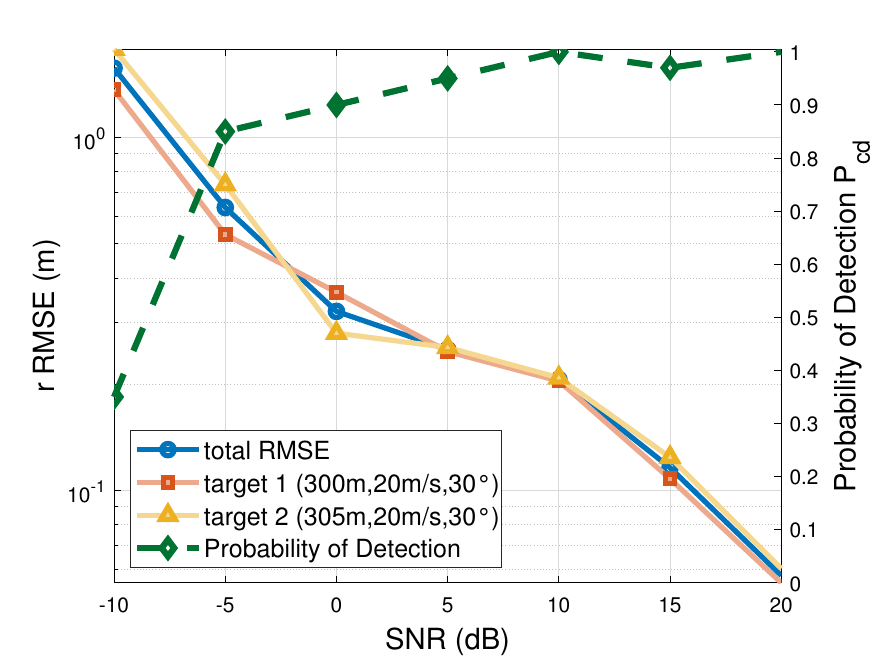}
		\captionsetup{font={small}}
		\caption{Range RMSE and probability of detection}
		\label{sr_range}
	\end{subfigure}
	\hfill
	\begin{subfigure}[t]{0.325\textwidth}
		\centering
		\includegraphics[width=\textwidth]{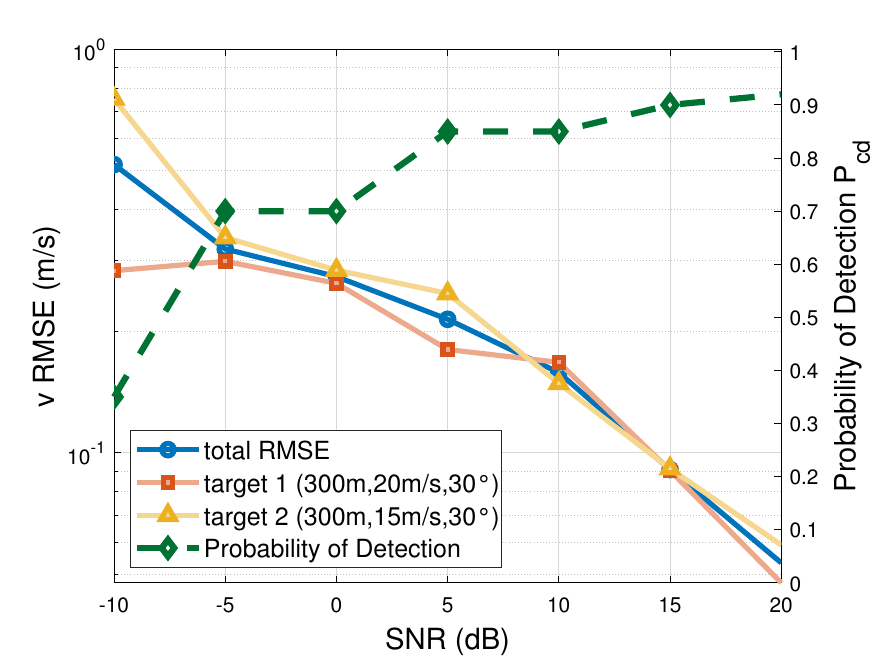} 
		\captionsetup{font={small}}
		\caption{Velocity RMSE and probability of detection}
		\label{sr_velocity}
	\end{subfigure}
	\hfill
	\begin{subfigure}[t]{0.32\textwidth}
		\centering
		\includegraphics[width=\textwidth]{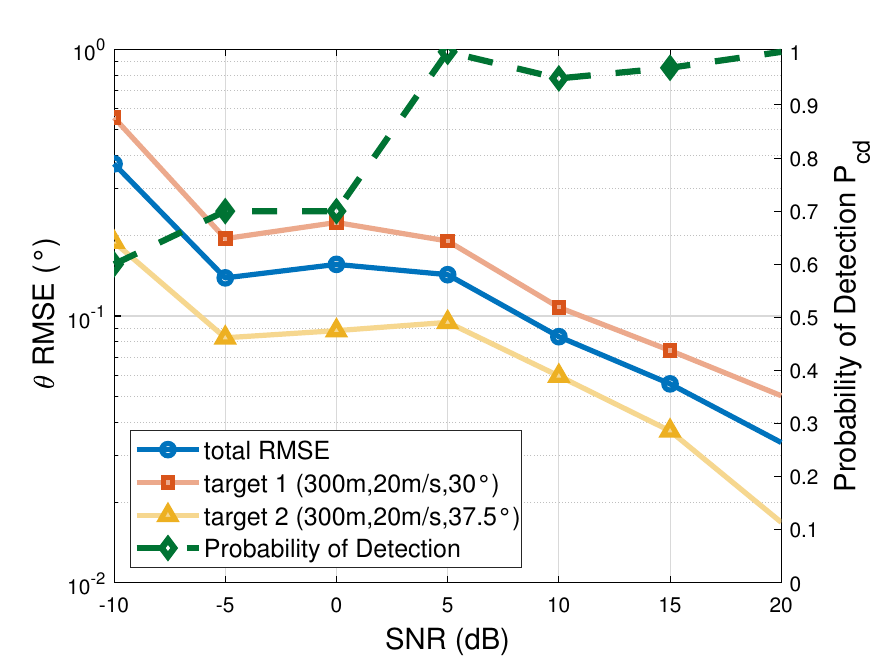}
		\captionsetup{font={small}}
		\caption{Angle RMSE and probability of detection}
		\label{sr_angle}
	\end{subfigure}
	\captionsetup{font={small}}
	\caption{Super-resolution performance in range/velocity/angle dimensions.}
	\label{sr}
	\vspace{-4mm}
\end{figure*}

\vspace{-2mm}
\subsection{Super-Resolution Performance}
Resolution is the performance limit of an ISAC system in distinguishing adjacent targets, which defines the theoretical upper bound of match filtering-based parameter estimation techniques. Based on our system configuration, the theoretical Rayleigh resolution for range, velocity, and angle are
\begin{equation}
	\begin{aligned}
		&\Delta r = \frac{c}{2M\Delta f}\approx 10\ \mathrm{m}, \\
		&\Delta v = \frac{\lambda}{2KT_{\mathrm{s}}} \approx 30\ \mathrm{m/s},\\
		&\Delta \theta = 0.886\frac{\lambda}{D} \approx  14.5^{\circ}.
	\end{aligned}
\end{equation}
Crucially, our continuous parameter space formulation breaks through these barriers, enabling reliable target differentiation below these theoretical limits.

Fig. \ref{sr} validates the algorithm's breakthrough super-resolution capabilities across all three dimensions by placing two targets at separations well below these limits.
As shown in Fig. ~\ref{sr}(\subref{sr_range}), with targets separated by just 5 m ($50\%$ below $\Delta r$), the algorithm successfully maintains a detection probability of over $80\%$ for SNRs above 0 dB. Furthermore, it achieves centimeter-level estimation accuracy for both targets with minimal bias between them, demonstrating fair and balanced performance.
Similarly, velocity and angle results in Fig. \ref{sr}(\subref{sr_velocity}) and Fig. \ref{sr}(\subref{sr_angle}) demonstrate robust super-resolution at 5 m/s ($83\%$ below $\Delta v$) and 7.5° ($48\%$ below $\Delta\theta$) separations with $> 75\%$ detection probability for SNR $>$ 0 dB with $< 3\%$ inter-target estimation bias, exceptional fairness between targets.
The framework's unified approach to super-resolution is evidenced by consistent detection probability trends across dimensions, characterized by rapid improvement from -10 dB to 0 dB followed by asymptotic convergence to near-perfect detection. 

Our algorithm breaks the Rayleigh resolution limit through continuous parameter space optimization, enabling target separation below theoretical bounds while ensuring balanced estimation across targets. Moreover, the multi-dimensional mutually enhances resolution that range-velocity-angle information fusion creates complementary enhancement where ambiguities in one dimension refine others. Unlike grid-constrained CS methods, our approach not only delivers superior resolution but also reduces the storage and computational complexity, making it suitable for the stringent performance and efficiency requirements in ISAC-LAWNs systems.

\vspace{-2mm}
\subsection{Clutter Suppression}
The preceding analyses are conducted in scenarios dominated by thermal noise. 
In practical ISAC LAWNs applications, the receiver's performance is often limited not by noise, but by strong environmental clutter.
This is particularly true in our mixed air-ground scenario, where undesired echoes from static or slow-moving ground objects can easily obscure the much weaker signals from dynamic targets, especially low-RCS UAVs. 
Clutter can be broadly categorized into two types: structured clutter from large, static, or slow-moving objects, which is characterized by high power and near-zero Doppler shift, and unstructured clutter from volumetric scattering, which is often treated as noise \cite{ieeeclutter}. The former poses the most significant challenge to target detection and estimation.

\vspace{-0.5mm}
In our simulations, we model this dominant structured clutter as echoes from $L_{\mathrm{C}}$ scatterers, where $L_{\mathrm{C}}$ is randomly drawn from the integer range [10,15]. These clutter paths share same distance and angle distributions as the dynamic targets, but their velocities are confined to a near-zero range of [-1,1] m/s. The path gain for clutter is modeled to be significantly stronger than that of a typical UAV target. To quantify the impact of this interference, we define the signal-to-clutter-and-noise ratio (SCNR) as
\begin{equation}
	\mathrm{SCNR} = \frac{P_{\mathrm{target}}}{P_{\mathrm{clutter}}+P_{\mathrm{noise}}}.
\end{equation}
To ensure the robustness of our proposed algorithm in such realistic scenarios, we employ a lightweight, real-time pre-processing module to suppress clutter before the core gridless SBL algorithm. We adopt a low-complexity background subtraction approach based on a computationally efficient low-pass filter that operates recursively across OFDM symbols \cite{PMN_cs}. This filter effectively separates the low-Doppler clutter from the target signals. After a brief transient phase of about 30 OFDM symbols ($\approx$ 0.5 ms), the filter's output stabilizes. To ensure maximum accuracy, the subsequent analysis for target detection and parameter estimation utilizes only this stable portion of the signal.

\begin{figure}[t]
	\vspace{-4mm}
	\centering
	\includegraphics[width=1\linewidth]{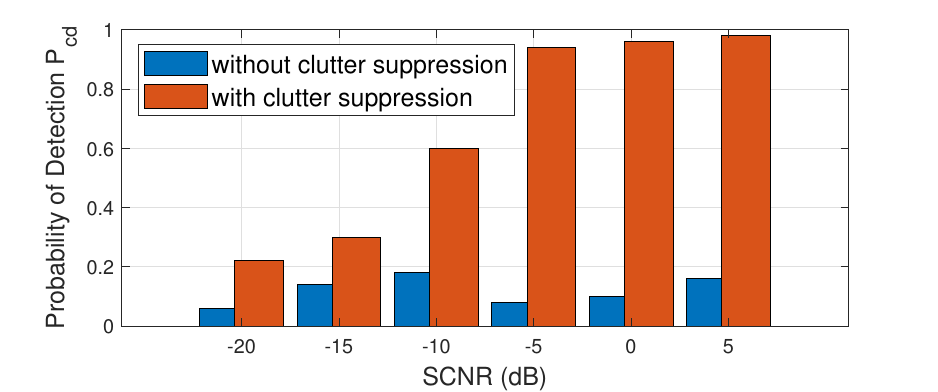}
	\captionsetup{font={small}}
	\caption{Probability of Detection versus SCNR.}
	\label{CP_Pcd}
\end{figure}

The effectiveness of this module is clearly demonstrated by the results. Fig. \ref{CP_Pcd} shows the detection rate as a function of SCNR. Without suppression, the detection rate is consistently below $20\%$, indicating that the target is overwhelmed by clutter. With suppression, the performance improves dramatically; for instance, at an SCNR of -10 dB, the detection rate jumps from under $20\%$ to $60\%$. This shows the module effectively turns a clutter-limited problem into a manageable noise-limited one.

\begin{figure}[t]
	\centering
	\vspace{-4mm}
	\includegraphics[width=0.85\linewidth]{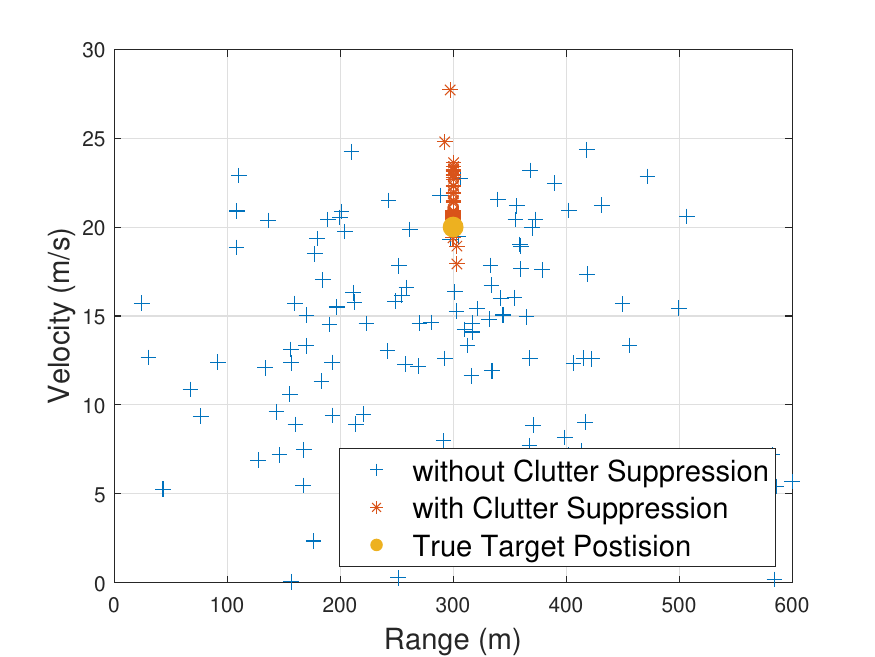}
	\captionsetup{font={small}}
	\caption{Range-Velocity estimation accuracy with/without suppression.}
	\label{clutter dot}
	\vspace{-4mm}
\end{figure}

Furthermore, clutter suppression significantly enhances parameter estimation accuracy, as visualized in Fig. \ref{clutter dot} from 50 Monte Carlo trials. Without suppression, the range-velocity estimates (blue crosses) are widely scattered and provide no reliable information. In contrast, after suppression, the estimates (orange asterisks) form a tight, well-defined cluster centered directly on the true target position (yellow circle), demonstrating a profound improvement in both precision and accuracy. This pre-processing step creates a much cleaner input for our gridless SBL algorithm, enabling it to maintain its high-accuracy performance even in the interference of strong clutter.

\section{Conclusion}
In this work, we proposed a novel MCMC-based gridless SBL framework for ISAC receivers to address the stringent sensing demands of LAWNs. By formulating the estimation task in a continuous parameter space, our gridless approach fundamentally eliminates the basis mismatch errors inherent in conventional grid-based methods, bypassing the need for large, discretized dictionaries. To make this powerful model computationally viable, we designed a highly efficient gradient-based MCMC sampler that integrates mini-batch sampling and the Adam optimizer, ensuring both rapid convergence and practical scalability.
Numerical results validate that our approach achieves both super-resolution performance that surpasses the Rayleigh limit and near-optimal estimation accuracy that closely tracks the derived BCRB. Furthermore, we demonstrated its robustness against strong ground clutter, showing its practical viability for realistic LAWN environments. Future work will explore adaptive MCMC schemes and hardware-efficient designs for real-time LAWNs deployments.

\vspace{-1mm}
\appendices  
\section{Derivation of Bayesian Cramér–Rao Bound}
This appendix provides the detailed derivation of BCRB for the real-valued parameter vector $\boldsymbol{\zeta}=[\mathbf{b}_{\mathrm{R}},\mathbf{b}_{\mathrm{I}},\boldsymbol{\tau},\boldsymbol{f}_{\mathrm{D}},\boldsymbol{\theta},\xi]^{\mathrm{T}}$. 
The core of the BCRB derivation lies in computing the BFIM. We first focus on deriving the classical FIM $\boldsymbol{\mathcal{I}}_{\mathrm{C}}$.
The echo signal at the resource element indexed by $r=(m,k)$ is expressed in a vectorized form using the Kronecker product identity as
\vspace{-1mm}
\begin{equation}
	\mathbf{y}_{r}=(\mathbf{x}_{r}^{\mathrm{T}}\otimes\mathbf{I})\times \mathbf{h}_{r}+\mathbf{z}_{r},  \quad \mathbf{I} \in \mathbb{R}^{N \times N},
	\label{echo_re}
	\vspace{-1mm}
\end{equation}
where $\mathbf{h}_{r} \in \mathbb{C}^{N^2}$ is the vectorized channel, given by
\vspace{-2mm}
\begin{equation}
	\mathbf{h}_{r}=\sum_{l=1}^{L_{\mathrm{T}}}b_l\omega_{r,l}\mathbf{a}(\theta_l)\otimes\mathbf{a}(\theta_l),
	\vspace{-1mm}
\end{equation}
where $\omega_{r,l}=e^{-j2\pi m \Delta f \tau_l}e^{j2\pi f_{\mathrm{D},l}kT_{\mathrm{s}}}$ for convenience. 
Given the echo signal is corrupted by complex additive Gaussian noise, the $(i, j)$-th entry of $\boldsymbol{\mathcal{I}}_{\mathrm{C}}$ is given by \cite{kay}
\begin{equation}
	[\boldsymbol{\mathcal { I }}_{\mathrm{C}}]_{i, j} \! = \! \mathrm{Tr}\left[\boldsymbol{C}_{\mathbf{y}}^{-1} \frac{\partial \boldsymbol{C}_{\mathbf{y}}}{\partial \zeta_i} \boldsymbol{C}_{\mathbf{y}}^{-1}  \frac{\partial \boldsymbol{C}_{\mathbf{y}}}{\partial \zeta_j} \right] 
	\! +\! 2 \mathrm{Re}\left[\frac{\partial \boldsymbol{\mu}^{\mathrm{H}}}{\partial \zeta_i}  \boldsymbol{C}_{\mathbf{y}}^{-1} \frac{\partial \boldsymbol{\mu}}{\partial \zeta_j} \right],
	\label{FIM}
\end{equation}
where $\boldsymbol{C}_{\mathbf{y}}=\xi^{-1} \mathbf{I}$ is variance matrix of posterior, and $\boldsymbol{\mu}=(\mathbf{x}_{m,k}^{\mathrm{T}}\otimes\mathbf{I})\times \mathbf{h}_{m,k}$ is the mean matrix in our model. 
The first trace term is non-zero only if both $\zeta_i$, $\zeta_j$ are $\xi$, while the second term is non-zero only for parameters in the mean $\boldsymbol{\mu}$. This decoupling leads to the block-diagonal structure of $\boldsymbol{\mathcal{I}}_{\mathrm{C}}$
\begin{equation}
	\boldsymbol{\mathcal{I}}_{\mathrm{C}}=
	\begin{bmatrix}
		\boldsymbol{\mathcal{I}}_{\mathrm{C}}^{s} & \mathbf{0} \\
		\mathbf{0} &  [\boldsymbol{\mathcal{I}}_{\mathrm{C}}]_{\xi,\xi}
		\label{diag_Ic}
	\end{bmatrix}.
\end{equation}
where $[\boldsymbol{\mathcal{I}}_{\mathrm{C}}]_{\xi,\xi}=MKN/\xi^{2}$.  $\boldsymbol{\mathcal{I}}_{\mathrm{C}}^{s}$ is the FIM block for all physical parameters except $\xi$, given by
\begin{equation}
	[\boldsymbol{\mathcal{I}}_{\mathrm{C}}]_{i,j}^{s}
	\!=\! \sum_{r=1}^{M \times K} \!2\xi \cdot \mathrm{Re}\left\{ \left(\frac{\partial\mathbf{h}_r}{\partial\zeta_i}\right)^{\mathrm{H}}\! (\mathbf{x}_r^*\mathbf{x}_r^{\mathrm{T}}  \! \otimes \! \mathbf{I})\left(\frac{\partial\mathbf{h}_r}{\partial\zeta_j}\right) \! \right\}.
\end{equation} 
The derivative of $\mathbf{h}_r$ to each parameter is
\begin{equation}
	\begin{aligned}
		&\frac{\partial\mathbf{h}_r}{\partial b_{l,\mathrm{R}}} =\omega_{r,l}\mathbf{a}(\theta_l)\otimes\mathbf{a}(\theta_l),\\
		&\frac{\partial\mathbf{h}_r}{\partial b_{l,\mathrm{I}}}
		=j\omega_{r,l}\mathbf{a}(\theta_l)\otimes\mathbf{a}(\theta_l),\\
		&\frac{\partial\mathbf{h}_r}{\partial \tau_{l}}
		= -j2\pi m \Delta f\times b_l \omega_{r,l} \mathbf{a}(\theta_l)\otimes\mathbf{a}(\theta_l), \\ 
		&\frac{\partial\mathbf{h}_r}{\partial f_{\mathrm{D},l}} 
		= j2\pi k T_{\mathrm{s}} \times b_l \omega_{r,l} \mathbf{a}(\theta_l)\otimes\mathbf{a}(\theta_l),\\
		&\frac{\partial\mathbf{h}_r}{\partial \theta_{l}}\!
		= \! b_l \omega_{r,l} \mathbf{D}_{\mathrm{sum}}(\theta_l) \times \big(\mathbf{a}(\theta_l)\otimes\mathbf{a}(\theta_l)\big),
	\end{aligned}
\end{equation}
where $\mathbf{D}_{\mathrm{sum}}(\theta_l) = \mathbf{D}_{\mathrm{array}}(\theta_l) \otimes \mathbf{I} + \mathbf{I} \otimes \mathbf{D}_{\mathrm{array}}(\theta_l)$, and
\begin{equation}
	[\mathbf{D}_{\mathrm{array}}(\theta_l)]_{nn} = j \frac{2\pi d n}{\lambda} \cos(\theta_l).
\end{equation}
These derivatives can be expressed in a compact form as
\begin{equation}
	\nabla_{\boldsymbol{\zeta}} \mathbf{h}_r = \left[ \frac{\partial \mathbf{h}_r}{\partial b_{l,\mathrm{R}}}, \cdots ,\frac{\partial \mathbf{h}_r}{\partial \theta_l} \right] = \mathbf{v}_r \cdot \mathbf{p}_r^{\mathrm{T}},
\end{equation} 
where $\mathbf{v}_r = \omega_{r,l}(\mathbf{a}(\theta_l) \otimes \mathbf{a}(\theta_l))$ and the modulation vector $\mathbf{p}_r$ is defined as 
\vspace{-1mm}
\begin{equation}
	\mathbf{p}_r = \left[1, \quad \! \!\!\!\! j,\quad \! \!\!\!\! b_l(-j2\pi m\Delta f),\quad \! \!\!\!\! b_l(j2\pi kT_{{\mathrm{s}}}), \quad \! \!\!\!\! b_l\frac{\mathbf{v}_r^{{\mathrm{H}}} \mathbf{D}_{\mathrm{sum}} \mathbf{v}_r}{\|\mathbf{v}_r\|^2} \right]^{\mathrm{T}},
\end{equation}
\vspace{-1mm}
the resulting $\boldsymbol{\mathcal{I}}_{\mathrm{C}}^{s}$ is derived as 
\begin{equation}
	\boldsymbol{\mathcal{I}}_{\mathrm{C}}^{s} = \sum_{r=1}^{M \times K} 2\xi \cdot \mathrm{Re}\left\{ \mathbf{p}_r^* \left( \mathbf{v}_r^{\mathrm{H}} (\mathbf{x}_r^* \mathbf{x}_r^{\mathrm{T}} \otimes \mathbf{I}) \mathbf{v}_r \right) \mathbf{p}_r^{\mathrm{T}} \right\}.
	\vspace{-2mm}
\end{equation}
The prior information matrix $\boldsymbol{\mathcal{I}}_{\mathrm{P}}=\mathbb{E}_{\boldsymbol{\zeta}}\left[-\frac{\partial^2 \log p(\boldsymbol{\zeta})}{\partial \boldsymbol{\zeta}^2} \right]$ is diagonal due to the independent priors.
With a hierarchical sparse prior, the real and imaginary components of $b_l$ are distributed as $\mathcal{N}(0,\rho_l/2)$, where the variance $\rho_l$ follows a Gamma distribution, the prior term of $b_{l,\mathrm{R}}$ is
\vspace{-1mm}
\begin{equation}
	\boldsymbol{\mathcal{I}}_{\mathrm{P}}(b_{l,\mathrm{R}}) = 
	\mathbb{E}_{\rho_l}\left[\frac{2}{\rho_l}\right]=\frac{2\chi_{\uprho}}{\kappa_{\uprho}-1},
	\vspace{-1mm}
\end{equation}
where the expectation over $\rho_l$ is valid under the condition $\kappa_{\uprho}>1$. The prior term for $b_{l,\mathrm{I}}$ is identical in form to $b_{l,\mathrm{R}}$.
For parameters $\boldsymbol{\tau}$, $\boldsymbol{f}_{\mathrm{D}}$, and $\boldsymbol{\theta}$ modeled with a truncated Gaussian prior, the prior information matrix can be computed as
\vspace{-1mm}
\begin{equation}
	[\boldsymbol{\mathcal{I}}_{\mathrm{P}}]_{\tau_l,\tau_l}=\mathbb{E}_{\tau_l}\left[\frac{1}{\sigma_{\uptau}^2}\right]=\frac{1}{\sigma_{\uptau}^2},
\end{equation}
where $\boldsymbol{f}_{\mathrm{D}}$ and $\boldsymbol{\theta}$ have the same form as $\boldsymbol{\mathcal{I}}_{\mathrm{P}}(\tau_l)$.
With a Gamma prior, the entry for $\xi$ is $[\boldsymbol{\mathcal{I}}_{\mathrm{P}}]_{\xi,\xi}=\chi_{\upxi}^2/(\kappa_{\upxi}-2)$.

Since the expectation $\mathbb{E}_{\boldsymbol{\zeta}}[\boldsymbol{\mathcal{I}}_{\mathrm{C}}]$ is analytically intractable, we approximate it via Monte Carlo integration over $S$ prior samples
\begin{equation}
	\vspace{-2mm}
	\hat{\boldsymbol{\mathcal{I}}}_{\mathrm{B}} \approx \frac{1}{S} \sum_{s=1}^S \boldsymbol{\mathcal{I}}_{\mathrm{C}}(\boldsymbol{\zeta}^{(s)})+\boldsymbol{\mathcal{I}}_{\mathrm{P}},
	\label{MC_BFIM}
\end{equation}
and the BCRB for each parameter can be obtained by (\ref{BCRB_def}).

\bibliography{ref}

\end{document}